\documentclass[%
 reprint,
 amsmath,amssymb,
 aps,
]{revtex4-2}

\usepackage{graphicx}
\usepackage{dcolumn}
\usepackage{bm}
\usepackage[unicode=true,colorlinks=true,citecolor=blue,urlcolor=blue]{hyperref}
\usepackage{gensymb}

\begin{document}

\preprint{APS/123-QED}

\title{Constrained tandem neural network assisted inverse design of metasurfaces for microwave absorption}

\author{Xiangxu, He}
\author{Xiaohan, Cui}%


\author{C. T. Chan}
\email[Corresponding author: ]{phchan@ust.hk}

\affiliation{Department of Physics, Hong Kong University of Science and Technology, Hong Kong, China}







\date{\today}

\begin{abstract}
Designing microwave absorbers with customized spectrums is an attractive topic in both scientific and engineering communities. However, due to the massive number of design parameters involved, the design process is typically time-consuming and computationally expensive. To address this challenge, machine learning has emerged as a powerful tool for optimizing design parameters. In this work, we present an analytical model for an absorber composed of a multi-layered metasurface and propose a novel inverse design method based on a constrained tandem neural network. The network can provide structural and material parameters optimized for a given absorption spectrum, without requiring professional knowledge. Furthermore, additional physical attributes, such as absorber thickness, can be optimized when soft constraints are applied. As an illustrative example, we use the neural network to design broadband microwave absorbers with a thickness close to the causality limit imposed by the Kramers-Kronig relation. Our approach provides new insights into the reverse engineering of physical devices.

\end{abstract}

\maketitle


\section{Introduction}
Electromagnetic absorbers have widespread use in various fields, including industrial sensors\cite{EMsensor1,EMsensor2}, wireless communication\cite{EMcomm1}, emitters\cite{EMemit1,EMemit2}, and thermophotovoltaics\cite{EMther1}. The Salisbury screen, which absorbs microwave effectively at a single frequency, served as the pioneering design\cite{SalScr1,SalScr2}. Subsequent advancements, such as the Jaumann absorber\cite{Jaum1,Jaum2} and circuit analog, have resulted in thinner and lighter microwave absorbers, with a broader working bandwidth within the targeted frequency range. However, small thickness and large working bandwidth are
conflicting requirements due to the limitation of Kramers-Kronig relations\cite{Cir1,KK1,KK2}. Recently, microwave absorbers based on metasurfaces have been proposed\cite{EMemit1,MetaAbso1,MetaAbso2,MetaAbso3,MetaAbso4,MetaAbso5,MetaAbso6,MetaAbso7,MetaAbso8}. These metamaterial absorbers have sub-wavelength geometric patterns that can induce multiple electric and magnetic resonances, providing numerous design parameters that can be tuned to satisfy diverse requirements. However, the relation between design parameters and electromagnetic response is often quite complicated, and tuning the design parameters to achieve the target response is effectively an optimization process in a high-dimensional parameter space, which is tedious and time-consuming. Indeed, it would be highly desirable if the design problem can be solved as an inverse scattering problem, which itself is very challenging for electromagnetic waves.

In recent years, machine learning has found applications in many areas. With the ability to learn from data, the machine learning system can find the hidden characteristics of the data and establish the relationship between the data sets. It has been used successfully in many fields, such as computer vision\cite{AIcv1,AIcv2,AIcv3,AIcv4,AIcv5}, natural language processing\cite{AInlp1,AInlp2,AInlp3,AInlp4,AInlp5}, time series forecasting\cite{AItsf1,AItsf2,AItsf3,AItsf4,AItsf5}, etc. In the field of physics, machine learning has also been used extensively\cite{AIphy1,AIphy2,AIphy3,AIphy4,AIphy5,AIphy6}, including solving the inverse problem\cite{AIinv1,AIinv2,AIinv3,AIinv4,AIinv5,AIinv6,AIinv7,AIinv8}. 
Recent studies have demonstrated the powerful capabilities of machine learning in the design of optical and nanophotonic devices\cite{rec1,rec2,rec3,rec4,rec5,rec6,rec7,rec9,rec10}. 
However, achieving a design with specific optimized physical attributes is challenging due to the extensive experimentation and iteration required to identify the optimal combination of physical parameters. This often involves the use of complex artificial neural networks (ANNs) and iterative processes.
Therefore, it is worthwhile to further explore and demonstrate how machine learning can assist in designing a microwave absorber to achieve a target absorption spectrum. Moreover, it would be even more desirable if the designed absorber could simultaneously have a thickness close to the causality limit, representing the thinnest achievable in passive media.

In this work, we demonstrate the potential of ANN in realizing the inverse design of a broadband microwave absorber based on metasurfaces. Using the training data generated from an analytical model of a microwave absorber composed of a multilayered metasurface, we establish and train a forward network that maps the design parameters of the absorber to its frequency response(in our case, it is the reflection spectrum). To overcome the significant challenge of one-to-many mapping in the inverse design\cite{AIinv1}, we build a tandem network by connecting the forward network to an inverse network. In addition, we included the constraint on the thickness of the design as a penalization term in the cost function to prevent the design from being excessively bulky.
After the training process, our constrained tandem neural network(CTNN) can not only provide us with a design that satisfies the target absorption spectrum but also has a thickness that is very close to the causality limit. 
Our investigation indicates that machine learning can effectively solve inverse design problems with specific requirements, and our approach can be readily extended to design acoustic and elastic wave absorbers.

\section{Generating Training Data}

We will first introduce the fundamental structure of the microwave absorbers that we aim to design. Metamaterial/metasurfaces, composed of a periodic array of subwavelength resonant elements, can manipulate the electric and magnetic response. By carefully tuning the resonant elements, good absorption can be achieved\cite{MetaAbso2}. Nevertheless, such a single-layer absorber typically has a narrow absorption band. To achieve broadband absorption, many multilayered metasurfaces with multiple resonant modes have been proposed\cite{LayAbso1,LayAbso2}. However, the increased thickness of such structures may render them too bulky for various applications. According to the Kramers-Kronig relations, for the normal incident illumination, there is a theoretical limit for the total thickness of a metal-backed non-magnetic absorber with relative permeability $\mu_r$=1 at the static limit\cite{Cir1,rozanov2000ultimate}
\begin{equation}
d_{\rm{tot}} \geq d_{\rm {min}} = \frac{\left|\int_{0}^{\infty} \ln | \Gamma(\lambda)|d \lambda\right|}{2 \pi^{2}},
\end{equation}
where $\Gamma(\lambda)$ is the reflection coefficient as a function of the wavelength $\lambda$.  This expression suggests that a high-performing absorber, capable of effectively absorbing a wide range of wavelengths, typically requires a considerable minimum thickness.
Therefore, designing a broadband absorber with a small thickness is challenging.    

\begin{figure*}[ht]
\centering
\includegraphics[width=\textwidth]{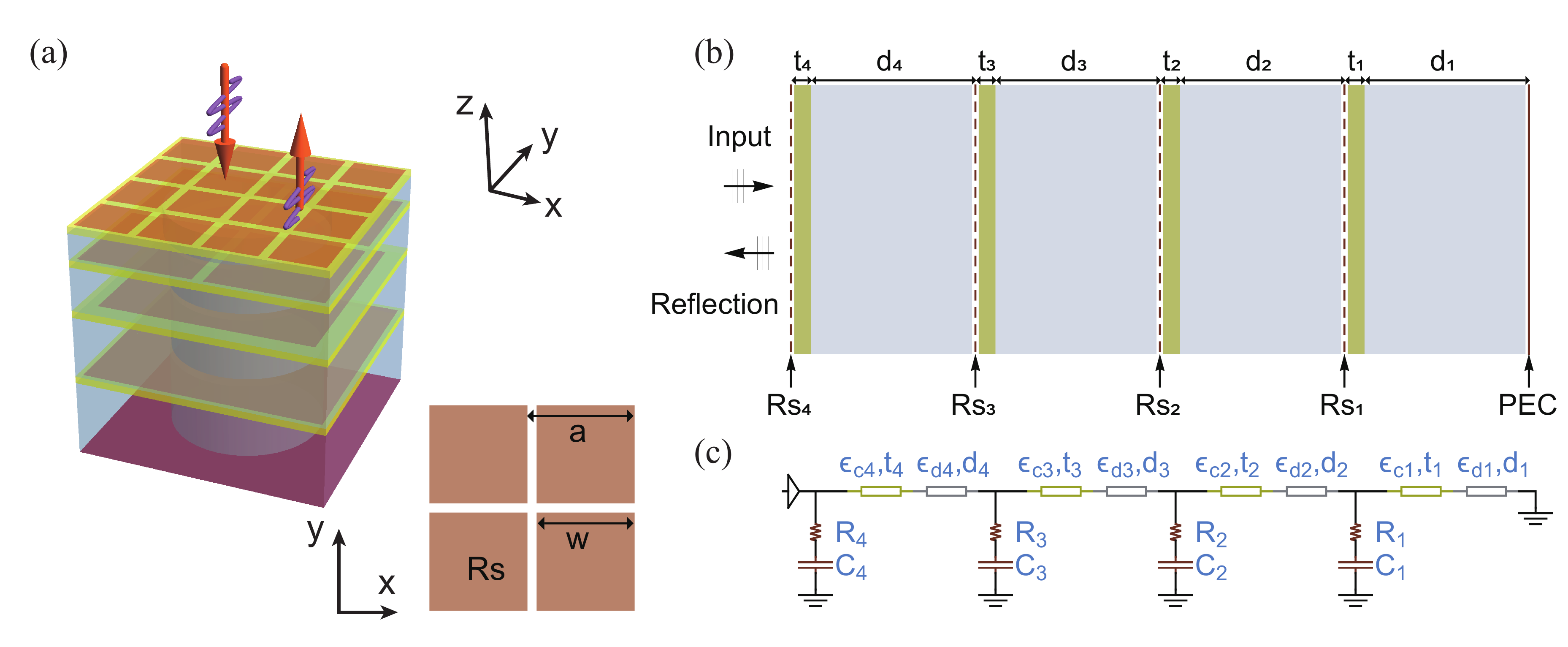}
\caption{Microwave absorber and its equivalent circuit model. (a) Schematic of a 4-layer microwave absorber composed of resistive square patch arrays (brown), dielectric substrates (yellow), and spacers (gray). A top-view picture of the patch array is shown at the lower right. $Rs$, $a$, and $w$ are the sheet resistance of the patch, the lattice constant, and the length of the square, respectively. (b) The side view of the absorber. $d_{1,2,3,4}$ and $t_{1,2,3,4}$ are the thickness of the spacers and substrates of different layers. (c) The equivalent circuit model of the absorber. The $i$th resistive square patch array can be modeled as a series RC circuit with resistance $R_i$ and capacitance $C_i$. The dielectric substrates and spacers can be represented by transmission line sections with lengths $t_i$ and $d_i$, respectively.}
\label{FIG1}
\end{figure*}

Our objective is to design a structure that possesses excellent absorption properties across a wide frequency range, while maintaining a thickness that approaches the minimum limit dictated by causality.
We consider a stacked multilayered structure as depicted in Fig. \ref{FIG1}. Such a structure is chosen because it can be fabricated in a layer-by-layer manner, and the electromagnetic response can be simulated using simple models. The proposed broadband absorber comprises four layers of metasurfaces (resistive patches placed on dielectric substrates (yellow)), which are separated by spacer layers (grey). In the $i$ th layer ($i=1,2,3,4$), the metasurface is a periodic array of resistive square patches (brown) with sheet resistance $Rs_i$, periodicity $a_i$, and square length $w_i$. 
This multilayered structure is placed on a flat metallic ground plane (red), which works as a perfect electric conductor (PEC), resulting in zero transmission. Therefore, the absorption of the multilayered structure is $1-|\Gamma|^2$, and the perfect absorption is achieved when $|\Gamma|=0$. This generally requires the absorber to be impedance-matched to the free space.

Although simulation software packages such as COMSOL can be used to obtain the absorption spectrum of arbitrary metamaterial structures numerically, the full-wave simulation of 3D problems is time-consuming, particularly when generating absorption spectra of multiple absorbers with varying parameters. Therefore, for this design, we utilized the simple structure shown in Fig. \ref{FIG1} that employs the square patch array, which enables us to solve the absorption spectrum almost analytically. By employing the capacitive circuit absorber approach\cite{CapaCir1}, we constructed an equivalent circuit model for this multilayered structure, as shown in Fig. \ref{FIG1}(c).
This enabled us to efficiently generate a large amount of training data for machine learning.

The capacitive circuit absorber approach uses the low-pass RC circuits instead of the conventional resonant RLC circuits to model the resistive patches with large square length ($w/a>0.7$)\cite{CapaCir1}.
For each layer, the resistive square patch can be modeled as a series RC circuit with resistance $Re_i$ and capacitance $C_i$, which are complex functions of the patch parameters ($Rs_i$, $a_i$, $w_i$). 
For the normal incidence, the values of the lumped elements can be evaluated analytically\cite{CapaCir2},
\begin{align}
    Re_{i}&=R s_{i} \frac{a_{i}^{2}}{w_{i}^{2}}\\
    C_{i}(\lambda)&=4 Y_{0} \varepsilon_{{\rm{r}} i} \frac{\lambda w_{i}}{2 \pi c a_{i}} F\left(a_{i}, a_{i}-w_{i}, \lambda\right),
\end{align}
where $Y_0$ is the admittance of the free space, and $c$ is the light of speed. $\epsilon_{{\rm r}i}=(\epsilon_{{\rm d}i}+\epsilon_{{\rm c}i})/2$ is the averaged relative permittivity of the spacers and the substrates in which the patch is embedded\cite{CapaCir3}. The function $F$ is\cite{RCmodel} 
\begin{widetext}
\begin{equation}
F(a, g, \lambda)=\frac{a}{\lambda}\left\{\ln \left(\operatorname{cosec} \frac{\pi g}{2 a}\right)+\frac{1}{2} \frac{\left(1-\alpha^{2}\right)^{2}\left[2 A\left(1-\frac{\alpha^{2}}{4}\right)+4 \alpha^{2} A^{2}\right]}{\left(1-\frac{\alpha^{2}}{4}\right)+2 A \alpha^{2}\left(1+\frac{\alpha^{2}}{2}-\frac{\alpha^{4}}{8}\right)+2 \alpha^{6} A^{2}}\right\},
\end{equation}
\end{widetext}
where $g=a-w$ represents the gap between patches, $A=\left[1-(a / \lambda)^{2}\right]^{-1 / 2}-1$ and $\alpha=\sin (\pi g / 2 a)$.Therefore, the admittance of the $i$th series RC circuit is\cite{CapaCir4} 
\begin{equation}
Y_{i}^{\rm s}=\left(Re_{i}+\frac{\lambda}{i 2 \pi c C_{i}}\right)^{-1}.
\end{equation}

The dielectric substrates and spacers can be represented by transmission line sections with lengths $t_i$ and $d_i$, respectively. According to the transmission line theory\cite{CapaCir4}, we can calculate the input admittance of different layers iteratively,
\begin{align}
Y_{i}&=Y_{i}^{\rm s}+Y_{{\rm c} i} \frac{\tilde{Y}_{i}+Y_{{\rm c} i} \tanh \left(i \beta_{{\rm c} i} t_{i}\right)}{Y_{{\rm c} i}+\tilde{Y}_{i} \tanh \left(i \beta_{{\rm c} i} t_{i}\right)}, \\
\tilde{Y}_{i}&=Y_{{\rm d} i} \frac{Y_{i-1}+Y_{{\rm d} i} \tanh \left(i \beta_{{\rm d} i} d_{i}\right)}{Y_{{\rm d} i}+Y_{i-1} \tanh \left(i \beta_{{\rm d} i} d_{i}\right)},
\end{align}
where $Y_{{\rm c} i}=\sqrt{\varepsilon_{{\rm c} i}} Y_{0}$, $Y_{{\rm d} i}=\sqrt{\varepsilon_{{\rm d} i}} Y_{0}$, $\beta_{{\rm c} i}=2 \pi \sqrt{\varepsilon_{{\rm c} i}} / \lambda,$ and $\beta_{{\rm d} i}=2 \pi \sqrt{\varepsilon_{{\rm d} i}} / \lambda$. The spacer of the first layer is backed by PEC and can be regarded as a short-circuited transmission line with the input admittance
\begin{equation}
\tilde{Y}_{1}=Y_{{\rm d} 1} \operatorname{coth}\left(i \beta_{{\rm d} 1} d_{1}\right).
\end{equation}

Therefore, for the transmission line circuit in Fig. \ref{FIG1}(b), the reflection coefficient is 
\begin{equation}\label{eq-Gamma}
\Gamma=\frac{Y_{0}-Y_{\rm in}}{Y_{0}+Y_{\rm in}},
\end{equation}
where $Y_{\rm in}$ represents the input admittance of the absorber and is equal to the input impedance of the fourth layer $Y_{\rm in}=Y_4$.
Using Eq.\eqref{eq-Gamma}, the reflection spectrum $R(\lambda)=20\log_{10}|\Gamma(\lambda)|$ of this 4-layer microwave absorber can be obtained analytically. As shown in Fig. \ref{FIG3}(b), the analytical results are in good agreement with the results obtained from the full-wave simulation.

\section{Constrained Tandem Neural Network}

In this part, we will introduce how to use the constrained tandem neural network to realize the inverse design of metasurface microwave absorbers that operates within a specified frequency range.
To represent the design parameters of the microwave absorber, we utilize an array $D=\{D_1, D_2, ... ,D_N\}$, and we refer to the array $R=\{R_1, R_2, ..., R_M\}$ as the response vector, which represents the reflection at discretely sampled frequency points within the specific range.
In this way, our objective can be reinterpreted as creating the inverse mapping from the response vector $R$ to the design parameter vector $D$, which is commonly referred to as an inverse problem.

It is natural to use an ANN to form this mapping directly, in view of the ability of such a network to fit highly-nonlinear functions\cite{UnivAppr}. However, this approach proves ineffective due to the presence of a one-to-many relationship within the dataset. 
There exist multiple design parameter vectors $D^{\rm{real}}$  which accord to the given response vector $R$. 
Therefore, the loss function of the network
\begin{equation}
E_{\rm {direct}}= \sum_{i}\left(D_{i}^{\rm {pred}}-D_{i}^{\rm {real}}\right)^{2}
\end{equation}
will encounter difficulties as the design parameter vector $D^{\rm {pred }}$ bounces around in parameter space when attempting to converge toward multiple target objectives represented by $D^{\rm {real}}$.  
Such a struggle often results in significant challenges or even outright failure in achieving network converge.

\begin{figure*}[ht]
\centering
\includegraphics[width=\textwidth]{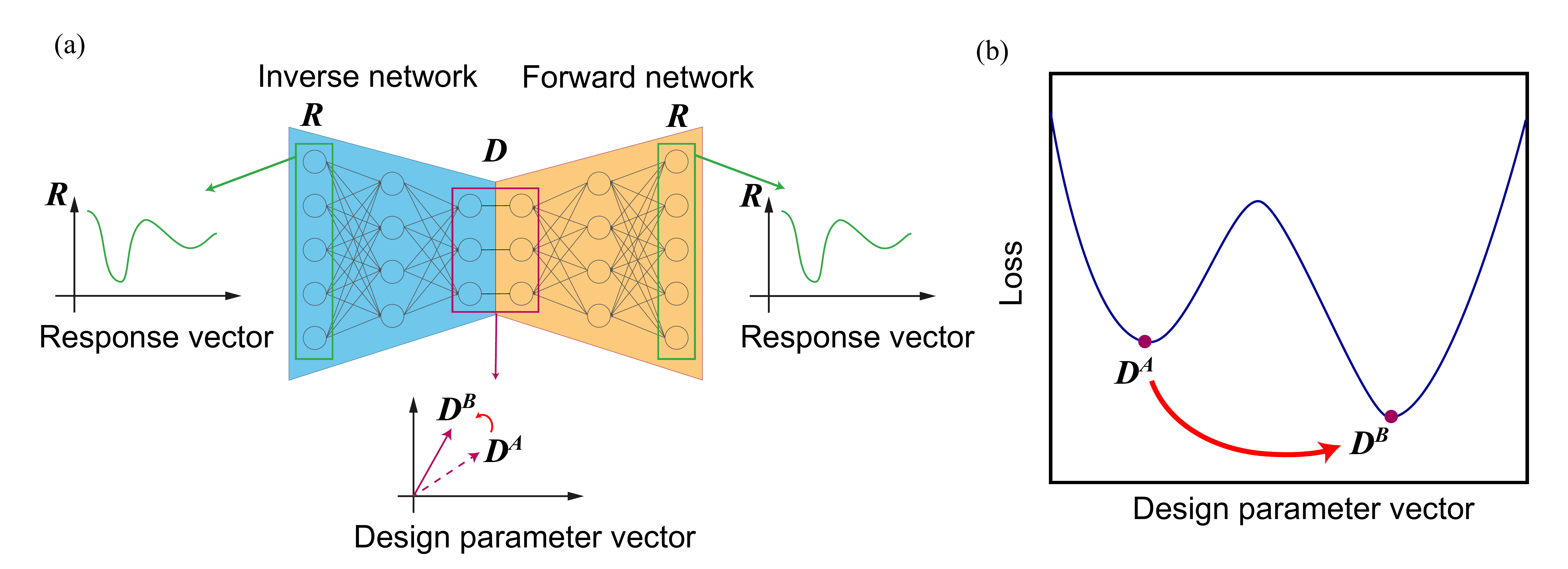}
\caption{
Constrained Tandem Neural Network(CTNN). (a) The CTNN is a tandem network composed of the forward network (orange) $D-R$ and the inverse network (blue) $R-D$. The forward network takes the latent vector $D$ as input and the corresponding response $R$ as output, while the inverse network takes $R$ as input and $D$ as output.
In our case, components of the array $D$ are exactly the design parameters, and $R$ represents reflectivity at discrete frequencies. 
A soft constraint imposed on the loss function can differentiate the designs $D^{\rm A}$ and $D^{\rm B}$ corresponding to the same response. (b) With the punishment of the thicker design, the latent vector will finally converge at a extreme point where  the design with thickness closer to the causality limit (as the red arrow shows).
}
\label{FIG2}
\end{figure*}

 To overcome this issue, the tandem neural network(TNN) method is proposed\cite{AIinv5}. TNN consists of two parts: the forward network (the orange part) and the inverse network (the blue part), shown in Fig. \ref{FIG2}(a). As we have stated above, the forward network takes $D$ (the right part of the crimson rectangle) as the input and $R$ (the right green rectangle) as the output, while the inverse network (the left green rectangle) takes the $R$ as the input and $D$ (the left part of the crimson rectangle) vector as the output.  The two parts of the tandem network are connected end-to-end, meaning that the output of the inverse network is the input of the forward network.
 Such $R$-$D$-$R$ structure is similar to the encoder-decoder ANN, and the intermediate output $D$ can be regarded as the latent vector, which encodes the information in the response vector $R$.
 Broadly speaking, the latent vector can be assigned with different physical meanings on different occasions. On some occasions, it just stands for "compressed information" without explicit physical meaning. 

To train the TNN, one should first train the forward network, enabling it to form the mapping from the $D$ vector to the response. After the training, the forward network should be frozen, which means that the weights and the bias of the forward network are kept unchanged in the subsequent steps. The TNN will then be trained to minimize the difference between the input of the inverse network and the output of the forward network. In this process, only the weights and the bias of the inverse network will be changed. 

As such, the loss function can be expressed as:
\begin{equation}
E_{\rm{TNN}}=\sum_{i}\left(R_i^{\rm {real }}-R_i^{\rm {pred}}\right)^{2}.
\end{equation}
Here, $R^{\rm{real}}$ means the ground truth corresponding to the design, while $R^{\rm{pred}}$ means the predicted response given by the network. 
Since the output R is unique, there will be no conflict arising from competing objectives. When the training of TNN is completed, the inverse mapping from $R$ to $D$ has been established.

However, there are also some drawbacks to this method. While the output $R$ of the TNN is unique, there are multiple possible choices for the intermediate output $D$. As the network is unbiased towards different $D$, the design parameters provided by the TNN have a significant degree of randomness. Therefore, it may provide us with design parameters that are undesirable, such as those with an excessively large thickness. 

To impose additional control over the intermediate output design parameters, we proposed the constrained tandem neural network (CTNN). Different from the ordinary TNN, we introduce a soft constraint term into the loss function, 
\begin{equation}
E_{\rm{CTNN}}=\sum_{i}(R_i^{\rm {real }}-R_i^{\rm {pred}})^{2} + \alpha (d_{\rm {tot}}-d_{\rm min})^{2},
\end{equation}
where $(d_{\rm {tot}}-d_{\rm min})^{2}$ represents the difference between the total thickness $d_{\rm{tot}}$ of the absorber and the minimum thickness $d_{\rm{min}}$, and $\alpha$ is a constant coefficient to control the second loss term. We have the flexibility to include multiple constraint terms as desired, but our primary focus is on a term that prioritizes the selection of thinner slabs as thickness is a critical factor when it comes to absorber design.    
The constraint incurs an additional cost to penalize thicker designs, thereby favoring thinner meta-structures that yield the same response. As Fig. \ref{FIG2}(b) illustrates pictorially that although there are two different design parameter vectors $D^A$ and $D^B$ that yield essentially the same response, the extra penalty term differentiates them because they have different total thicknesses. 
And this penalty will lead the CTNN to converge towards $D^B$, because the total thickness of the design $D^B$ is closer to the causality limit than $D^A$, $(d_{\rm{tot}}^B-d_{\rm{min}})^2<(d_{\rm{tot}}^A-d_{\rm{min}})^2$. In this way, the arbitrariness of the design parameter vector $D$ is removed. 

Additionally, since the soft constraint term can be chosen as desired, it enables control over other design parameters, making it possible to meet various design requirements. For example, by introducing a constraint related to the sheet resistance Rs, we can realize a design with smaller resistance values if we choose to do so, ensuring that the material parameters can be constrained to a pre-specified range. We also note that this method can be generalized to the inverse design for other systems since the physical meaning of the latent vector and the constraint can be defined according to the real occasion.

\section{Network Training}

In the training process of CTNN, the steps are the same as those in TNN. The only change is the customized loss function incorporating constraint. Therefore, the training time is almost the same as TNN. After the training, the end-to-end application is enabled and no more training is needed.

Considering the fabrication feasibility, we simplify the design by requiring the parameters of the dielectric substrate of different layers to be the same value $\epsilon_{{\rm{c}} i}=\epsilon_{\rm {C}}$, $t_i=T_{\rm {C}}$, and fixing the dielectric constant of the spacer as $\epsilon_{{\rm{d}}i}=\epsilon_{\rm {d}}=1.2$, which can be realized by using porous PVC substrates. The lattice constant $a_i = 6.8$ mm, $6.8$ mm, $3.4$ mm, $1.7$ mm for $i = 1,2,3,4$, respectively. These structural parameters have been used previously to design absorbers\cite{Cir1}. The parameters used for inverse design are the length of the partches($w_{1,2,3,4}$), sheet resistance ($Rs_{1,2,3,4}$), thickness($d_{1,2,3,4}$) of the spacers, dielectric constant($\epsilon_{\rm {C}}$) and thickness($T_{\rm {C}}$) of the substrate, forming a 14$\times$1 design parameter vector $D=\{\epsilon_{\rm {C}}, T_{\rm {C}}, d_1, d_2, d_3, d_4, w_1, w_2, w_3, w_4, Rs_1, Rs_2, Rs_3, Rs_4\}$. And the total thickness of the absorber is $d_{\rm{tot}}=4T_{\rm{C}}+\sum_{i=1}^4 d_i$.
The reflectivity $|\Gamma(f)|^2$ is measured at 320 evenly distributed discrete frequencies $f_i\in[1$ GHz,$41$ GHz$]$, $i=1,2,...,320$, forming a 320$\times$1 response vector $R$. 
Using the circuit model in Fig. \ref{FIG1}(c), we can generate training samples analytically.

The forward network $D-R$ has 4 hidden layers with 80, 320, 320, and 320 neurons for different layers respectively to avoid sudden zooming, and the activation function leaky ReLU($k=0.01$) is adopted. The input and output layer contains 14 and 320 units respectively, to match the data. We take $\alpha = 0.01$ in the constraint term. 
The inverse network $R-D$ also contains 4 hidden layers with 320, 320, 320, and 80 neurons respectively, and the Adam optimizer is adopted with a learning rate of $10^{-4}$.

\begin{figure*}[ht]
\centering
\includegraphics[width=\textwidth]{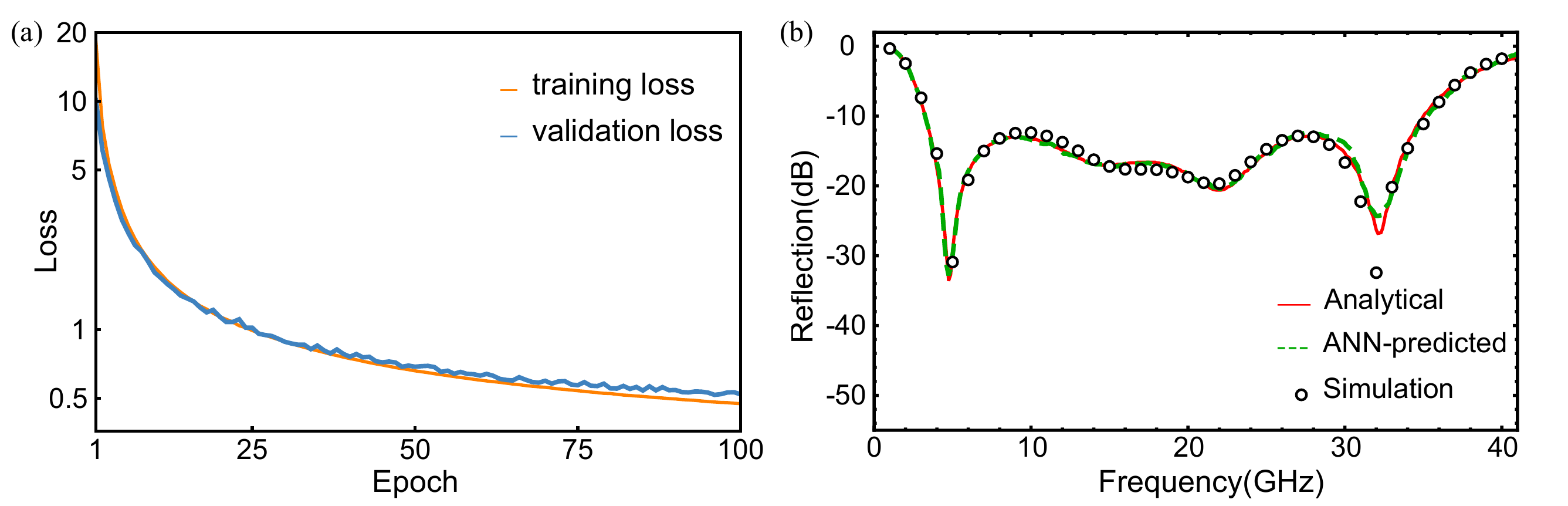}
\caption{Forward training of data. (a) Loss descending during the training. The orange line denotes the training loss while the blue line denotes the validation loss. After training for 100 epochs, the forward network achieved convergence. (b) The absorption response obtained by the analytical model, forward network $D - R$, and full-wave numerical simulation are represented by the red line, green dashed line, and open circles, respectively.
}
\label{FIG3}
\end{figure*}

Within a pre-specified range of parameters(specified in the 2nd column in Table \ref{table1}), we randomly generate the training data. We use $10^5$ $D-R$ supervised data to train the forward network. 
As shown in Fig.\ref{FIG3}(a), after 100 epochs of training of the forward network, the loss decreases to an acceptable value. The design parameters and the corresponding response of one of the test samples are shown in Table. \ref{table1} and Fig. \ref{FIG3}(b).
Then, we use $10^5$ $R$ data to train CTNN. The data is distributed by training: valuation: testing = 8: 1: 1. Figure \ref{FIG4} (a) demonstrates that after 50 epochs of training, the loss function decreases to an acceptable value. In the following section, we will provide examples of utilizing this well-trained CTNN for designing microwave absorbers.

\begin{table*}[ht]
\centering
\caption{The design parameters of the absorber. The random sampling region of design parameters in the training sets is given in the 2nd column. The sample from the test set shown in Fig. \ref{FIG3}(b) is given in the 3rd column, and two specific absorber designs given respectively by CTNN and TNN corresponding to our desired absorption responses shown in the Fig. \ref{FIG4} (b) are given in the 4th and 5th column.}
\begin{ruledtabular}
\begin{tabular}{  p{2.5cm}  p{2.5cm}  p{2.5cm}  p{2.5cm}  p{2.5cm}}
Parameter & Training range & Fig. \ref{FIG3}(b) & Fig. \ref{FIG4}(b) CTNN & Fig. \ref{FIG4}(b) TNN \\
\hline
$\epsilon_S$        & -           & 1.20             & 1.20            & 1.20            \\
$\epsilon_C$         & (2, 6)      & 4.95             & 6.32          & 7.64          \\
$T_C$(mm)    & (0.1, 1)    & 0.240             & 0.0553        & 0.0191          \\
$d_1$(mm)    & (1.8, 4.5)  & 3.27             & 3.03          & 3.39          \\
$d_2$(mm)    & (1.8, 3.5)  & 3.03             & 3.25          & 3.23          \\
$d_3$(mm)    & (1.8, 3.8)  & 2.63             & 3.15          & 3.48          \\
$d_4$(mm)    & (1.8, 3.8)  & 2.25             & 2.91          & 2.83          \\
$w_1$(mm)    & (6.0, 6.8)  & 6.69             & 6.79            & 6.50            \\
$w_2$(mm)    & (6.0, 6.5)  & 6.38             & 6.09            & 5.77            \\
$w_3$(mm)    & (2.7, 3.35) & 2.73            & 2.87            & 3.10            \\
$w_4$(mm)    & (1.2, 1.65) & 1.28             & 1.11            & 0.955            \\
$a_1$(mm)    & -           & 6.80             & 6.80            & 6.80            \\
$a_2$(mm)    & -           & 6.80             & 6.80            & 6.80            \\
$a_3$(mm)    & -           & 3.40             & 3.40            & 3.40            \\
$a_4$(mm)    & -           & 1.70             & 1.70            & 1.70            \\
$Rs_1$($\Omega$/Sq) & (20, 300)   & 240             & 167          & 172          \\
$Rs_2$($\Omega$/Sq) & (60, 700)   & 588             & 313          & 337          \\
$Rs_3$($\Omega$/Sq) & (300, 700)  & 512             & 472          & 603          \\
$Rs_4$($\Omega$/Sq) & (500, 1000) & 812             & 613           & 472      \\
\end{tabular}
\label{table1}
\end{ruledtabular}
\end{table*}

\section{Inverse Design of Microwave Absorbers}
\begin{figure*}[ht]
\centering
\includegraphics[width=\textwidth]{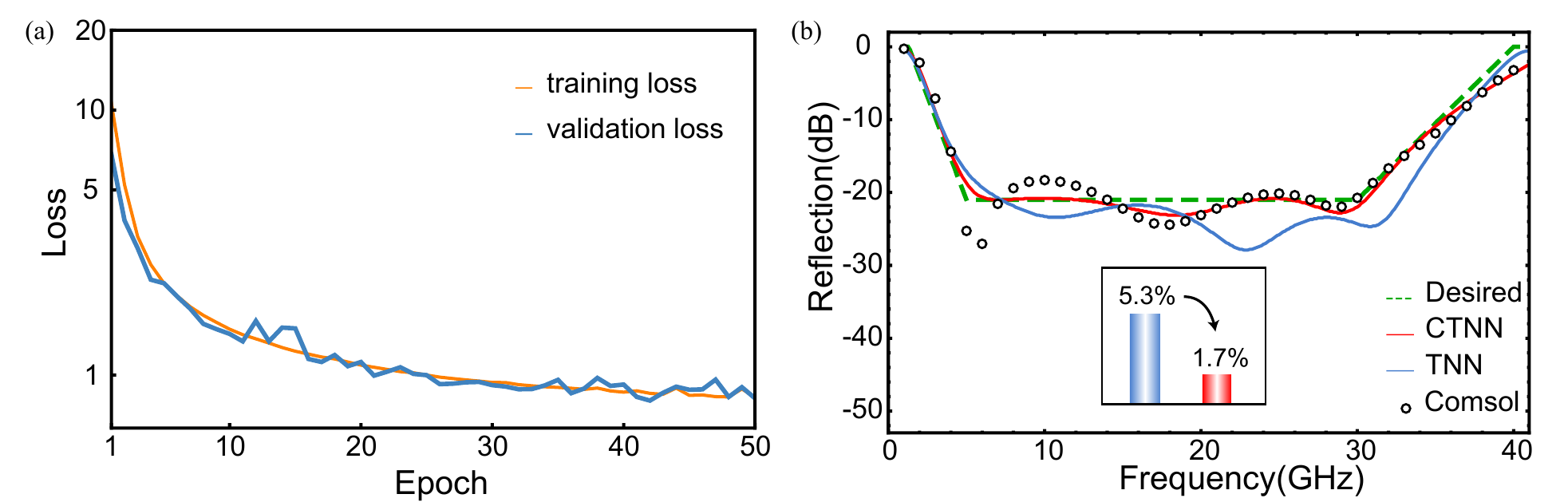}
\caption{Training of CTNN and the design of the broadband absorber. (a) Loss descending during the training. The orange line denotes the training loss while the blue line denotes the validation loss. (b) Using the CTNN to design the absorber with the desired absorption response denoted by dashed green lines. 
The absorption response obtained by the improved CTNN, the traditional TNN, and the full-wave numerical simulation are represented by the red line, blue line, and open circles, respectively.
Compared to the traditional TNN (with a thickness $5.3$\% thicker than the causality limit), the improved CTNN (with thickness $1.7$\% thicker than the causality limit) exhibits better performance. 
}
\label{FIG4}
\end{figure*}

Based on the trained neural network, we can now design the EM wave absorber according to a given absorption spectrum. Here we show a specific example in Fig. \ref{FIG4}(a), where we want to design a broadband absorber capable of achieving a 20 dB absorption across the frequency range of $4.5$ to $31.5$ GHz. Mathematically, the input response has a trapezoid-like shape (the green dashed line):
\begin{equation}
|\Gamma(f)|^2=\left\{\begin{array}{cc}
-\frac{C_{0}\left(f-f_{1}\right)}{f_{2}-f_{1}} & f_{1} \leq f \leq f_{2} \\
-C_{0} & f_{2} \leq f \leq f_{3} \\
\frac{C_{0}\left(f-f_{4}\right)}{f_{4}-f_{3}} & f_{3} \leq f \leq f_{4}
\end{array}\right..
\end{equation}
Here we set $C_0=21$ dB, $f_1=1.3$ GHz, $f_2=5$ GHz, $f_3=30$ GHz, $f_4=40$ GHz. Notice that we provide the desired spectrum using reflection coefficients. We input the desired spectrum into the trained tandem network. The intermediate layer output gives the design parameters, as shown in the 4th column of Table. \ref{table1}. In Fig. \ref{FIG4}, the desired absorption spectra, as specified mathematically by Eq. (5), is shown as the green dashed line. The red line in Fig. \ref{FIG4} is the absorption spectra of this designed structure calculated the analytical method while the black open dots are the absorption calculated for the same structure using full wave computation (using the package COMSOL). 

These results show that absorber we designed has absorption characteristics that match well with the required absorption spectra, which is shown as the green dashed line. In particular, the absorber has a $20$ dB reflection reduction covering the frequency interval $4.5$ to $31.5$ GHz. We also calculated the minimum absorber thickness corresponding to our desired spectrum as required by causality, which is found to be $12.339$ mm. The thickness of our design is $12.552$ mm, only 1.7\% thicker than the causality limit.  For comparison, the ordinary TNN gives a 5.3\% thicker design(shown in the 5th column of Table. \ref{table1}). More designs featuring various response spectra are included in the supplementary materials to illustrate the versatility of our network. We should also note that the design parameters are experimentally realizable. The required value of $\epsilon_C$ can be realized by various plastics like PVC and PET and the ITO films can serve as the resistive square patches\cite{ITO}.

\section{Conclusion}
In conclusion, we demonstrated a CTNN-assisted approach to design a custom microwave absorber that meets a pre-described frequency response requirement. This approach utilizes the tandem network to solve one-to-many problems with great effectiveness and efficiency. Furthermore, we added a soft constraint on the loss function to enable the network to achieve biased convergence towards a specific target, namely minimal thickness in this case. We construct the relation between absorber design vector $D$ and its absorption response vector $R$ by training the forward and the tandem network, respectively. After the training, our network can conduct the inverse design and give corresponding design parameters based on input-customized absorption responses. We use the network to design a broadband absorber, and the thickness is very close to the causality limit. We note that data from numerical simulations or experiments can also replace the training data for the network if such information is available. The approach can be easily applied to the inverse design of optical, acoustic, and other devices.

\begin{acknowledgments}
We acknowledges the support of the Hong Kong Research Grants Council's Research Impact Fund R6015-18 for this work. We would like to thank Prof. Ping Sheng, Prof. Wing Yim Tam, Prof. Zhaoqing Zhang, Dr. Kun Ding for useful discussions. X. Cui also wishes to thank Mr. Yufan Gao for the preliminary simulation work.
\end{acknowledgments}

\nocite{*}

\bibliography{apssamp}

\providecommand{\noopsort}[1]{}\providecommand{\singleletter}[1]{#1}%
\begin{thebibliography}{76}%
\makeatletter
\providecommand \@ifxundefined [1]{%
 \@ifx{#1\undefined}
}%
\providecommand \@ifnum [1]{%
 \ifnum #1\expandafter \@firstoftwo
 \else \expandafter \@secondoftwo
 \fi
}%
\providecommand \@ifx [1]{%
 \ifx #1\expandafter \@firstoftwo
 \else \expandafter \@secondoftwo
 \fi
}%
\providecommand \natexlab [1]{#1}%
\providecommand \enquote  [1]{``#1''}%
\providecommand \bibnamefont  [1]{#1}%
\providecommand \bibfnamefont [1]{#1}%
\providecommand \citenamefont [1]{#1}%
\providecommand \href@noop [0]{\@secondoftwo}%
\providecommand \href [0]{\begingroup \@sanitize@url \@href}%
\providecommand \@href[1]{\@@startlink{#1}\@@href}%
\providecommand \@@href[1]{\endgroup#1\@@endlink}%
\providecommand \@sanitize@url [0]{\catcode `\\12\catcode `\$12\catcode `\&12\catcode `\#12\catcode `\^12\catcode `\_12\catcode `\%12\relax}%
\providecommand \@@startlink[1]{}%
\providecommand \@@endlink[0]{}%
\providecommand \url  [0]{\begingroup\@sanitize@url \@url }%
\providecommand \@url [1]{\endgroup\@href {#1}{\urlprefix }}%
\providecommand \urlprefix  [0]{URL }%
\providecommand \Eprint [0]{\href }%
\providecommand \doibase [0]{https://doi.org/}%
\providecommand \selectlanguage [0]{\@gobble}%
\providecommand \bibinfo  [0]{\@secondoftwo}%
\providecommand \bibfield  [0]{\@secondoftwo}%
\providecommand \translation [1]{[#1]}%
\providecommand \BibitemOpen [0]{}%
\providecommand \bibitemStop [0]{}%
\providecommand \bibitemNoStop [0]{.\EOS\space}%
\providecommand \EOS [0]{\spacefactor3000\relax}%
\providecommand \BibitemShut  [1]{\csname bibitem#1\endcsname}%
\let\auto@bib@innerbib\@empty
\bibitem [{\citenamefont {Cheng}\ \emph {et~al.}(2016)\citenamefont {Cheng}, \citenamefont {Mao}, \citenamefont {Wu}, \citenamefont {Wu},\ and\ \citenamefont {Gong}}]{EMsensor1}%
  \BibitemOpen
  \bibfield  {author} {\bibinfo {author} {\bibfnamefont {Y.}~\bibnamefont {Cheng}}, \bibinfo {author} {\bibfnamefont {X.~S.}\ \bibnamefont {Mao}}, \bibinfo {author} {\bibfnamefont {C.}~\bibnamefont {Wu}}, \bibinfo {author} {\bibfnamefont {L.}~\bibnamefont {Wu}},\ and\ \bibinfo {author} {\bibfnamefont {R.}~\bibnamefont {Gong}},\ }\bibfield  {title} {\bibinfo {title} {Infrared non-planar plasmonic perfect absorber for enhanced sensitive refractive index sensing},\ }\href {https://doi.org/https://doi.org/10.1016/j.optmat.2016.01.053} {\bibfield  {journal} {\bibinfo  {journal} {Optical Materials}\ }\textbf {\bibinfo {volume} {53}},\ \bibinfo {pages} {195} (\bibinfo {year} {2016})}\BibitemShut {NoStop}%
\bibitem [{\citenamefont {Unal}\ \emph {et~al.}(2015)\citenamefont {Unal}, \citenamefont {Dincer}, \citenamefont {Tetik}, \citenamefont {Karaaslan}, \citenamefont {Bakir},\ and\ \citenamefont {Sabah}}]{EMsensor2}%
  \BibitemOpen
  \bibfield  {author} {\bibinfo {author} {\bibfnamefont {E.}~\bibnamefont {Unal}}, \bibinfo {author} {\bibfnamefont {F.}~\bibnamefont {Dincer}}, \bibinfo {author} {\bibfnamefont {E.}~\bibnamefont {Tetik}}, \bibinfo {author} {\bibfnamefont {M.}~\bibnamefont {Karaaslan}}, \bibinfo {author} {\bibfnamefont {M.}~\bibnamefont {Bakir}},\ and\ \bibinfo {author} {\bibfnamefont {C.}~\bibnamefont {Sabah}},\ }\bibfield  {title} {\bibinfo {title} {Tunable perfect metamaterial absorber design using the golden ratio and energy harvesting and sensor applications},\ }\href@noop {} {\bibfield  {journal} {\bibinfo  {journal} {Journal of Materials Science: Materials in Electronics}\ }\textbf {\bibinfo {volume} {26}},\ \bibinfo {pages} {9735} (\bibinfo {year} {2015})}\BibitemShut {NoStop}%
\bibitem [{\citenamefont {Namai}\ \emph {et~al.}(2009)\citenamefont {Namai}, \citenamefont {Sakurai}, \citenamefont {Nakajima}, \citenamefont {Suemoto}, \citenamefont {Matsumoto}, \citenamefont {Goto}, \citenamefont {Sasaki},\ and\ \citenamefont {Ohkoshi}}]{EMcomm1}%
  \BibitemOpen
  \bibfield  {author} {\bibinfo {author} {\bibfnamefont {A.}~\bibnamefont {Namai}}, \bibinfo {author} {\bibfnamefont {S.}~\bibnamefont {Sakurai}}, \bibinfo {author} {\bibfnamefont {M.}~\bibnamefont {Nakajima}}, \bibinfo {author} {\bibfnamefont {T.}~\bibnamefont {Suemoto}}, \bibinfo {author} {\bibfnamefont {K.}~\bibnamefont {Matsumoto}}, \bibinfo {author} {\bibfnamefont {M.}~\bibnamefont {Goto}}, \bibinfo {author} {\bibfnamefont {S.}~\bibnamefont {Sasaki}},\ and\ \bibinfo {author} {\bibfnamefont {S.-i.}\ \bibnamefont {Ohkoshi}},\ }\bibfield  {title} {\bibinfo {title} {Synthesis of an electromagnetic wave absorber for high-speed wireless communication},\ }\href {https://doi.org/10.1021/ja807943v} {\bibfield  {journal} {\bibinfo  {journal} {Journal of the American Chemical Society}\ }\textbf {\bibinfo {volume} {131}},\ \bibinfo {pages} {1170} (\bibinfo {year} {2009})}\BibitemShut {NoStop}%
\bibitem [{\citenamefont {Watts}\ \emph {et~al.}(2012)\citenamefont {Watts}, \citenamefont {Liu},\ and\ \citenamefont {Padilla}}]{EMemit1}%
  \BibitemOpen
  \bibfield  {author} {\bibinfo {author} {\bibfnamefont {C.~M.}\ \bibnamefont {Watts}}, \bibinfo {author} {\bibfnamefont {X.}~\bibnamefont {Liu}},\ and\ \bibinfo {author} {\bibfnamefont {W.~J.}\ \bibnamefont {Padilla}},\ }\bibfield  {title} {\bibinfo {title} {Metamaterial electromagnetic wave absorbers},\ }\href@noop {} {\bibfield  {journal} {\bibinfo  {journal} {Advanced Materials}\ }\textbf {\bibinfo {volume} {24}},\ \bibinfo {pages} {OP98} (\bibinfo {year} {2012})}\BibitemShut {NoStop}%
\bibitem [{\citenamefont {Cui}\ \emph {et~al.}(2014)\citenamefont {Cui}, \citenamefont {He}, \citenamefont {Jin}, \citenamefont {Ding}, \citenamefont {Yang}, \citenamefont {Ye}, \citenamefont {Zhong}, \citenamefont {Lin},\ and\ \citenamefont {He}}]{EMemit2}%
  \BibitemOpen
  \bibfield  {author} {\bibinfo {author} {\bibfnamefont {Y.}~\bibnamefont {Cui}}, \bibinfo {author} {\bibfnamefont {Y.}~\bibnamefont {He}}, \bibinfo {author} {\bibfnamefont {Y.}~\bibnamefont {Jin}}, \bibinfo {author} {\bibfnamefont {F.}~\bibnamefont {Ding}}, \bibinfo {author} {\bibfnamefont {L.}~\bibnamefont {Yang}}, \bibinfo {author} {\bibfnamefont {Y.}~\bibnamefont {Ye}}, \bibinfo {author} {\bibfnamefont {S.}~\bibnamefont {Zhong}}, \bibinfo {author} {\bibfnamefont {Y.}~\bibnamefont {Lin}},\ and\ \bibinfo {author} {\bibfnamefont {S.}~\bibnamefont {He}},\ }\bibfield  {title} {\bibinfo {title} {Plasmonic and metamaterial structures as electromagnetic absorbers},\ }\href@noop {} {\bibfield  {journal} {\bibinfo  {journal} {Laser Photonics Reviews}\ }\textbf {\bibinfo {volume} {8}},\ \bibinfo {pages} {495} (\bibinfo {year} {2014})}\BibitemShut {NoStop}%
\bibitem [{\citenamefont {Niu}\ \emph {et~al.}(2018)\citenamefont {Niu}, \citenamefont {Qi}, \citenamefont {Wang}, \citenamefont {Cheng}, \citenamefont {Chen}, \citenamefont {Li},\ and\ \citenamefont {Gong}}]{EMther1}%
  \BibitemOpen
  \bibfield  {author} {\bibinfo {author} {\bibfnamefont {X.}~\bibnamefont {Niu}}, \bibinfo {author} {\bibfnamefont {D.}~\bibnamefont {Qi}}, \bibinfo {author} {\bibfnamefont {X.}~\bibnamefont {Wang}}, \bibinfo {author} {\bibfnamefont {Y.}~\bibnamefont {Cheng}}, \bibinfo {author} {\bibfnamefont {F.}~\bibnamefont {Chen}}, \bibinfo {author} {\bibfnamefont {B.}~\bibnamefont {Li}},\ and\ \bibinfo {author} {\bibfnamefont {R.}~\bibnamefont {Gong}},\ }\bibfield  {title} {\bibinfo {title} {Improved broadband spectral selectivity of absorbers/emitters for solar thermophotovoltaics based on 2d photonic crystal heterostructures},\ }\href@noop {} {\bibfield  {journal} {\bibinfo  {journal} {J. Opt. Soc. Am. A}\ }\textbf {\bibinfo {volume} {35}},\ \bibinfo {pages} {1832} (\bibinfo {year} {2018})}\BibitemShut {NoStop}%
\bibitem [{\citenamefont {Salisbury}(1952)}]{SalScr1}%
  \BibitemOpen
  \bibfield  {author} {\bibinfo {author} {\bibfnamefont {W.~W.}\ \bibnamefont {Salisbury}},\ }\href@noop {} {\bibinfo {title} {Absorbent body for electromagnetic waves}} (\bibinfo {year} {U.S. Patent, 2,599,944, 1952})\BibitemShut {NoStop}%
\bibitem [{\citenamefont {Fante}\ and\ \citenamefont {McCormack}(1988)}]{SalScr2}%
  \BibitemOpen
  \bibfield  {author} {\bibinfo {author} {\bibfnamefont {R.}~\bibnamefont {Fante}}\ and\ \bibinfo {author} {\bibfnamefont {M.}~\bibnamefont {McCormack}},\ }\bibfield  {title} {\bibinfo {title} {Reflection properties of the salisbury screen},\ }\href {https://doi.org/10.1109/8.8632} {\bibfield  {journal} {\bibinfo  {journal} {IEEE Transactions on Antennas and Propagation}\ }\textbf {\bibinfo {volume} {36}},\ \bibinfo {pages} {1443} (\bibinfo {year} {1988})}\BibitemShut {NoStop}%
\bibitem [{\citenamefont {Knott}\ and\ \citenamefont {Lunden}(1995)}]{Jaum1}%
  \BibitemOpen
  \bibfield  {author} {\bibinfo {author} {\bibfnamefont {E.}~\bibnamefont {Knott}}\ and\ \bibinfo {author} {\bibfnamefont {C.}~\bibnamefont {Lunden}},\ }\bibfield  {title} {\bibinfo {title} {The two-sheet capacitive jaumann absorber},\ }\href {https://doi.org/10.1109/8.475112} {\bibfield  {journal} {\bibinfo  {journal} {IEEE Transactions on Antennas and Propagation}\ }\textbf {\bibinfo {volume} {43}},\ \bibinfo {pages} {1339} (\bibinfo {year} {1995})}\BibitemShut {NoStop}%
\bibitem [{\citenamefont {Du~Toit}(1994)}]{Jaum2}%
  \BibitemOpen
  \bibfield  {author} {\bibinfo {author} {\bibfnamefont {L.}~\bibnamefont {Du~Toit}},\ }\bibfield  {title} {\bibinfo {title} {The design of jauman absorbers},\ }\href {https://doi.org/10.1109/74.370526} {\bibfield  {journal} {\bibinfo  {journal} {IEEE Antennas and Propagation Magazine}\ }\textbf {\bibinfo {volume} {36}},\ \bibinfo {pages} {17} (\bibinfo {year} {1994})}\BibitemShut {NoStop}%
\bibitem [{\citenamefont {Kazemzadeh}(2011)}]{Cir1}%
  \BibitemOpen
  \bibfield  {author} {\bibinfo {author} {\bibfnamefont {A.}~\bibnamefont {Kazemzadeh}},\ }\bibfield  {title} {\bibinfo {title} {Nonmagnetic ultrawideband absorber with optimal thickness},\ }\href {https://doi.org/10.1109/TAP.2010.2090481} {\bibfield  {journal} {\bibinfo  {journal} {IEEE Transactions on Antennas and Propagation}\ }\textbf {\bibinfo {volume} {59}},\ \bibinfo {pages} {135} (\bibinfo {year} {2011})}\BibitemShut {NoStop}%
\bibitem [{\citenamefont {Jackson}(2007)}]{KK1}%
  \BibitemOpen
  \bibfield  {author} {\bibinfo {author} {\bibfnamefont {J.~D.}\ \bibnamefont {Jackson}},\ }\href@noop {} {\emph {\bibinfo {title} {Classical electrodynamics}}}\ (\bibinfo  {publisher} {John Wiley \& Sons},\ \bibinfo {year} {2007})\BibitemShut {NoStop}%
\bibitem [{\citenamefont {Nussenzveig}(1972)}]{KK2}%
  \BibitemOpen
  \bibfield  {author} {\bibinfo {author} {\bibfnamefont {H.~M.}\ \bibnamefont {Nussenzveig}},\ }\href@noop {} {\emph {\bibinfo {title} {Causality and dispersion relations}}}\ (\bibinfo  {publisher} {Academic Press},\ \bibinfo {year} {1972})\BibitemShut {NoStop}%
\bibitem [{\citenamefont {Luo}(2015)}]{MetaAbso1}%
  \BibitemOpen
  \bibfield  {author} {\bibinfo {author} {\bibfnamefont {X.}~\bibnamefont {Luo}},\ }\bibfield  {title} {\bibinfo {title} {Principles of electromagnetic waves in metasurfaces},\ }\href {https://doi.org/10.1007/s11433-015-5688-1} {\bibfield  {journal} {\bibinfo  {journal} {Science China Physics, Mechanics Astronomy}\ }\textbf {\bibinfo {volume} {58}},\ \bibinfo {pages} {594201} (\bibinfo {year} {2015})}\BibitemShut {NoStop}%
\bibitem [{\citenamefont {Landy}\ \emph {et~al.}(2008)\citenamefont {Landy}, \citenamefont {Sajuyigbe}, \citenamefont {Mock}, \citenamefont {Smith},\ and\ \citenamefont {Padilla}}]{MetaAbso2}%
  \BibitemOpen
  \bibfield  {author} {\bibinfo {author} {\bibfnamefont {N.~I.}\ \bibnamefont {Landy}}, \bibinfo {author} {\bibfnamefont {S.}~\bibnamefont {Sajuyigbe}}, \bibinfo {author} {\bibfnamefont {J.~J.}\ \bibnamefont {Mock}}, \bibinfo {author} {\bibfnamefont {D.~R.}\ \bibnamefont {Smith}},\ and\ \bibinfo {author} {\bibfnamefont {W.~J.}\ \bibnamefont {Padilla}},\ }\bibfield  {title} {\bibinfo {title} {Perfect metamaterial absorber},\ }\href {https://doi.org/10.1103/PhysRevLett.100.207402} {\bibfield  {journal} {\bibinfo  {journal} {Phys. Rev. Lett.}\ }\textbf {\bibinfo {volume} {100}},\ \bibinfo {pages} {207402} (\bibinfo {year} {2008})}\BibitemShut {NoStop}%
\bibitem [{\citenamefont {Liu}\ \emph {et~al.}(2010{\natexlab{a}})\citenamefont {Liu}, \citenamefont {Mesch}, \citenamefont {Weiss}, \citenamefont {Hentschel},\ and\ \citenamefont {Giessen}}]{MetaAbso3}%
  \BibitemOpen
  \bibfield  {author} {\bibinfo {author} {\bibfnamefont {N.}~\bibnamefont {Liu}}, \bibinfo {author} {\bibfnamefont {M.}~\bibnamefont {Mesch}}, \bibinfo {author} {\bibfnamefont {T.}~\bibnamefont {Weiss}}, \bibinfo {author} {\bibfnamefont {M.}~\bibnamefont {Hentschel}},\ and\ \bibinfo {author} {\bibfnamefont {H.}~\bibnamefont {Giessen}},\ }\bibfield  {title} {\bibinfo {title} {Infrared perfect absorber and its application as plasmonic sensor},\ }\href {https://doi.org/10.1021/nl9041033} {\bibfield  {journal} {\bibinfo  {journal} {Nano Letters}\ }\textbf {\bibinfo {volume} {10}},\ \bibinfo {pages} {2342} (\bibinfo {year} {2010}{\natexlab{a}})}\BibitemShut {NoStop}%
\bibitem [{\citenamefont {Liu}\ \emph {et~al.}(2010{\natexlab{b}})\citenamefont {Liu}, \citenamefont {Starr}, \citenamefont {Starr},\ and\ \citenamefont {Padilla}}]{MetaAbso4}%
  \BibitemOpen
  \bibfield  {author} {\bibinfo {author} {\bibfnamefont {X.}~\bibnamefont {Liu}}, \bibinfo {author} {\bibfnamefont {T.}~\bibnamefont {Starr}}, \bibinfo {author} {\bibfnamefont {A.~F.}\ \bibnamefont {Starr}},\ and\ \bibinfo {author} {\bibfnamefont {W.~J.}\ \bibnamefont {Padilla}},\ }\bibfield  {title} {\bibinfo {title} {Infrared spatial and frequency selective metamaterial with near-unity absorbance},\ }\href {https://doi.org/10.1103/PhysRevLett.104.207403} {\bibfield  {journal} {\bibinfo  {journal} {Phys. Rev. Lett.}\ }\textbf {\bibinfo {volume} {104}},\ \bibinfo {pages} {207403} (\bibinfo {year} {2010}{\natexlab{b}})}\BibitemShut {NoStop}%
\bibitem [{\citenamefont {Wang}\ \emph {et~al.}(2019)\citenamefont {Wang}, \citenamefont {Hou},\ and\ \citenamefont {Chan}}]{MetaAbso5}%
  \BibitemOpen
  \bibfield  {author} {\bibinfo {author} {\bibfnamefont {S.}~\bibnamefont {Wang}}, \bibinfo {author} {\bibfnamefont {B.}~\bibnamefont {Hou}},\ and\ \bibinfo {author} {\bibfnamefont {C.~T.}\ \bibnamefont {Chan}},\ }\bibfield  {title} {\bibinfo {title} {Broadband microwave absorption by logarithmic spiral metasurface},\ }\href@noop {} {\bibfield  {journal} {\bibinfo  {journal} {Scientific Reports}\ }\textbf {\bibinfo {volume} {9}},\ \bibinfo {pages} {1} (\bibinfo {year} {2019})}\BibitemShut {NoStop}%
\bibitem [{\citenamefont {Fan}\ \emph {et~al.}(2015)\citenamefont {Fan}, \citenamefont {Shen}, \citenamefont {Koschny},\ and\ \citenamefont {Soukoulis}}]{MetaAbso6}%
  \BibitemOpen
  \bibfield  {author} {\bibinfo {author} {\bibfnamefont {Y.}~\bibnamefont {Fan}}, \bibinfo {author} {\bibfnamefont {N.-H.}\ \bibnamefont {Shen}}, \bibinfo {author} {\bibfnamefont {T.}~\bibnamefont {Koschny}},\ and\ \bibinfo {author} {\bibfnamefont {C.~M.}\ \bibnamefont {Soukoulis}},\ }\bibfield  {title} {\bibinfo {title} {Tunable terahertz meta-surface with graphene cut-wires},\ }\href@noop {} {\bibfield  {journal} {\bibinfo  {journal} {ACS Photonics}\ }\textbf {\bibinfo {volume} {2}},\ \bibinfo {pages} {151} (\bibinfo {year} {2015})}\BibitemShut {NoStop}%
\bibitem [{\citenamefont {Zhang}\ and\ \citenamefont {Song}(2021)}]{MetaAbso7}%
  \BibitemOpen
  \bibfield  {author} {\bibinfo {author} {\bibfnamefont {M.}~\bibnamefont {Zhang}}\ and\ \bibinfo {author} {\bibfnamefont {Z.}~\bibnamefont {Song}},\ }\bibfield  {title} {\bibinfo {title} {Switchable terahertz metamaterial absorber with broadband absorption and multiband absorption},\ }\href@noop {} {\bibfield  {journal} {\bibinfo  {journal} {Opt. Express}\ }\textbf {\bibinfo {volume} {29}},\ \bibinfo {pages} {21551} (\bibinfo {year} {2021})}\BibitemShut {NoStop}%
\bibitem [{\citenamefont {Liu}\ \emph {et~al.}(2021)\citenamefont {Liu}, \citenamefont {Xu},\ and\ \citenamefont {Song}}]{MetaAbso8}%
  \BibitemOpen
  \bibfield  {author} {\bibinfo {author} {\bibfnamefont {W.}~\bibnamefont {Liu}}, \bibinfo {author} {\bibfnamefont {J.}~\bibnamefont {Xu}},\ and\ \bibinfo {author} {\bibfnamefont {Z.}~\bibnamefont {Song}},\ }\bibfield  {title} {\bibinfo {title} {Bifunctional terahertz modulator for beam steering and broadband absorption based on a hybrid structure of graphene and vanadium dioxide},\ }\href {https://doi.org/10.1364/OE.433364} {\bibfield  {journal} {\bibinfo  {journal} {Opt. Express}\ }\textbf {\bibinfo {volume} {29}},\ \bibinfo {pages} {23331} (\bibinfo {year} {2021})}\BibitemShut {NoStop}%
\bibitem [{\citenamefont {Sebe}\ \emph {et~al.}(2005)\citenamefont {Sebe}, \citenamefont {Cohen}, \citenamefont {Garg},\ and\ \citenamefont {Huang}}]{AIcv1}%
  \BibitemOpen
  \bibfield  {author} {\bibinfo {author} {\bibfnamefont {N.}~\bibnamefont {Sebe}}, \bibinfo {author} {\bibfnamefont {I.}~\bibnamefont {Cohen}}, \bibinfo {author} {\bibfnamefont {A.}~\bibnamefont {Garg}},\ and\ \bibinfo {author} {\bibfnamefont {T.~S.}\ \bibnamefont {Huang}},\ }\href@noop {} {\emph {\bibinfo {title} {Machine learning in computer vision}}},\ Vol.~\bibinfo {volume} {29}\ (\bibinfo  {publisher} {Springer Science \& Business Media},\ \bibinfo {year} {2005})\BibitemShut {NoStop}%
\bibitem [{\citenamefont {Szeliski}(2022)}]{AIcv2}%
  \BibitemOpen
  \bibfield  {author} {\bibinfo {author} {\bibfnamefont {R.}~\bibnamefont {Szeliski}},\ }\href@noop {} {\emph {\bibinfo {title} {Computer vision: algorithms and applications}}}\ (\bibinfo  {publisher} {Springer Nature},\ \bibinfo {year} {2022})\BibitemShut {NoStop}%
\bibitem [{\citenamefont {Guo}\ \emph {et~al.}(2022)\citenamefont {Guo}, \citenamefont {Xu}, \citenamefont {Liu}, \citenamefont {Liu}, \citenamefont {Jiang}, \citenamefont {Mu}, \citenamefont {Zhang}, \citenamefont {Martin}, \citenamefont {Cheng},\ and\ \citenamefont {Hu}}]{AIcv3}%
  \BibitemOpen
  \bibfield  {author} {\bibinfo {author} {\bibfnamefont {M.-H.}\ \bibnamefont {Guo}}, \bibinfo {author} {\bibfnamefont {T.-X.}\ \bibnamefont {Xu}}, \bibinfo {author} {\bibfnamefont {J.-J.}\ \bibnamefont {Liu}}, \bibinfo {author} {\bibfnamefont {Z.-N.}\ \bibnamefont {Liu}}, \bibinfo {author} {\bibfnamefont {P.-T.}\ \bibnamefont {Jiang}}, \bibinfo {author} {\bibfnamefont {T.-J.}\ \bibnamefont {Mu}}, \bibinfo {author} {\bibfnamefont {S.-H.}\ \bibnamefont {Zhang}}, \bibinfo {author} {\bibfnamefont {R.~R.}\ \bibnamefont {Martin}}, \bibinfo {author} {\bibfnamefont {M.-M.}\ \bibnamefont {Cheng}},\ and\ \bibinfo {author} {\bibfnamefont {S.-M.}\ \bibnamefont {Hu}},\ }\bibfield  {title} {\bibinfo {title} {Attention mechanisms in computer vision: A survey},\ }\href@noop {} {\bibfield  {journal} {\bibinfo  {journal} {Computational Visual Media}\ }\textbf {\bibinfo {volume} {8}},\ \bibinfo {pages} {331} (\bibinfo {year} {2022})}\BibitemShut {NoStop}%
\bibitem [{\citenamefont {Zhou}\ \emph {et~al.}(2023)\citenamefont {Zhou}, \citenamefont {Zhang},\ and\ \citenamefont {Konz}}]{AIcv4}%
  \BibitemOpen
  \bibfield  {author} {\bibinfo {author} {\bibfnamefont {L.}~\bibnamefont {Zhou}}, \bibinfo {author} {\bibfnamefont {L.}~\bibnamefont {Zhang}},\ and\ \bibinfo {author} {\bibfnamefont {N.}~\bibnamefont {Konz}},\ }\bibfield  {title} {\bibinfo {title} {Computer vision techniques in manufacturing},\ }\href {https://doi.org/10.1109/TSMC.2022.3166397} {\bibfield  {journal} {\bibinfo  {journal} {IEEE Transactions on Systems, Man, and Cybernetics: Systems}\ }\textbf {\bibinfo {volume} {53}},\ \bibinfo {pages} {105} (\bibinfo {year} {2023})}\BibitemShut {NoStop}%
\bibitem [{\citenamefont {Parvaiz}\ \emph {et~al.}(2023)\citenamefont {Parvaiz}, \citenamefont {Khalid}, \citenamefont {Zafar}, \citenamefont {Ameer}, \citenamefont {Ali},\ and\ \citenamefont {Fraz}}]{AIcv5}%
  \BibitemOpen
  \bibfield  {author} {\bibinfo {author} {\bibfnamefont {A.}~\bibnamefont {Parvaiz}}, \bibinfo {author} {\bibfnamefont {M.~A.}\ \bibnamefont {Khalid}}, \bibinfo {author} {\bibfnamefont {R.}~\bibnamefont {Zafar}}, \bibinfo {author} {\bibfnamefont {H.}~\bibnamefont {Ameer}}, \bibinfo {author} {\bibfnamefont {M.}~\bibnamefont {Ali}},\ and\ \bibinfo {author} {\bibfnamefont {M.~M.}\ \bibnamefont {Fraz}},\ }\bibfield  {title} {\bibinfo {title} {Vision transformers in medical computer vision—a contemplative retrospection},\ }\href@noop {} {\bibfield  {journal} {\bibinfo  {journal} {Engineering Applications of Artificial Intelligence}\ }\textbf {\bibinfo {volume} {122}},\ \bibinfo {pages} {106126} (\bibinfo {year} {2023})}\BibitemShut {NoStop}%
\bibitem [{\citenamefont {Chowdhary}(2020)}]{AInlp1}%
  \BibitemOpen
  \bibfield  {author} {\bibinfo {author} {\bibfnamefont {K.~R.}\ \bibnamefont {Chowdhary}},\ }\bibinfo {title} {Natural language processing},\ in\ \href@noop {} {\emph {\bibinfo {booktitle} {Fundamentals of Artificial Intelligence}}}\ (\bibinfo  {publisher} {Springer India},\ \bibinfo {year} {2020})\ pp.\ \bibinfo {pages} {603--649}\BibitemShut {NoStop}%
\bibitem [{\citenamefont {Sidorov}\ \emph {et~al.}(2014)\citenamefont {Sidorov}, \citenamefont {Velasquez}, \citenamefont {Stamatatos}, \citenamefont {Gelbukh},\ and\ \citenamefont {Chanona-Hern{\'a}ndez}}]{AInlp2}%
  \BibitemOpen
  \bibfield  {author} {\bibinfo {author} {\bibfnamefont {G.}~\bibnamefont {Sidorov}}, \bibinfo {author} {\bibfnamefont {F.}~\bibnamefont {Velasquez}}, \bibinfo {author} {\bibfnamefont {E.}~\bibnamefont {Stamatatos}}, \bibinfo {author} {\bibfnamefont {A.}~\bibnamefont {Gelbukh}},\ and\ \bibinfo {author} {\bibfnamefont {L.}~\bibnamefont {Chanona-Hern{\'a}ndez}},\ }\bibfield  {title} {\bibinfo {title} {Syntactic n-grams as machine learning features for natural language processing},\ }\href@noop {} {\bibfield  {journal} {\bibinfo  {journal} {Expert Systems with Applications}\ }\textbf {\bibinfo {volume} {41}},\ \bibinfo {pages} {853} (\bibinfo {year} {2014})}\BibitemShut {NoStop}%
\bibitem [{\citenamefont {Khurana}\ \emph {et~al.}(2023)\citenamefont {Khurana}, \citenamefont {Koli}, \citenamefont {Khatter},\ and\ \citenamefont {Singh}}]{AInlp3}%
  \BibitemOpen
  \bibfield  {author} {\bibinfo {author} {\bibfnamefont {D.}~\bibnamefont {Khurana}}, \bibinfo {author} {\bibfnamefont {A.}~\bibnamefont {Koli}}, \bibinfo {author} {\bibfnamefont {K.}~\bibnamefont {Khatter}},\ and\ \bibinfo {author} {\bibfnamefont {S.}~\bibnamefont {Singh}},\ }\bibfield  {title} {\bibinfo {title} {Natural language processing: state of the art, current trends and challenges},\ }\href {https://doi.org/10.1007/s11042-022-13428-4} {\bibfield  {journal} {\bibinfo  {journal} {Multimedia Tools and Applications}\ }\textbf {\bibinfo {volume} {82}},\ \bibinfo {pages} {3713} (\bibinfo {year} {2023})}\BibitemShut {NoStop}%
\bibitem [{\citenamefont {Wu}\ \emph {et~al.}(2023)\citenamefont {Wu}, \citenamefont {Chen}, \citenamefont {Shen}, \citenamefont {Guo}, \citenamefont {Gao}, \citenamefont {Li}, \citenamefont {Pei},\ and\ \citenamefont {Long}}]{AInlp4}%
  \BibitemOpen
  \bibfield  {author} {\bibinfo {author} {\bibfnamefont {L.}~\bibnamefont {Wu}}, \bibinfo {author} {\bibfnamefont {Y.}~\bibnamefont {Chen}}, \bibinfo {author} {\bibfnamefont {K.}~\bibnamefont {Shen}}, \bibinfo {author} {\bibfnamefont {X.}~\bibnamefont {Guo}}, \bibinfo {author} {\bibfnamefont {H.}~\bibnamefont {Gao}}, \bibinfo {author} {\bibfnamefont {S.}~\bibnamefont {Li}}, \bibinfo {author} {\bibfnamefont {J.}~\bibnamefont {Pei}},\ and\ \bibinfo {author} {\bibfnamefont {B.}~\bibnamefont {Long}},\ }\bibfield  {title} {\bibinfo {title} {Graph neural networks for natural language processing: A survey},\ }\href {https://doi.org/10.1561/2200000096} {\bibfield  {journal} {\bibinfo  {journal} {Foundations and Trends in Machine Learning}\ }\textbf {\bibinfo {volume} {16}},\ \bibinfo {pages} {119} (\bibinfo {year} {2023})}\BibitemShut {NoStop}%
\bibitem [{\citenamefont {Liu}\ \emph {et~al.}(2023)\citenamefont {Liu}, \citenamefont {Lin},\ and\ \citenamefont {Sun}}]{AInlp5}%
  \BibitemOpen
  \bibfield  {author} {\bibinfo {author} {\bibfnamefont {Z.}~\bibnamefont {Liu}}, \bibinfo {author} {\bibfnamefont {Y.}~\bibnamefont {Lin}},\ and\ \bibinfo {author} {\bibfnamefont {M.}~\bibnamefont {Sun}},\ }\href@noop {} {\emph {\bibinfo {title} {Representation learning for natural language processing}}}\ (\bibinfo  {publisher} {Springer Nature},\ \bibinfo {year} {2023})\BibitemShut {NoStop}%
\bibitem [{\citenamefont {Ahmed}\ \emph {et~al.}(2010)\citenamefont {Ahmed}, \citenamefont {Atiya}, \citenamefont {Gayar},\ and\ \citenamefont {El-Shishiny}}]{AItsf1}%
  \BibitemOpen
  \bibfield  {author} {\bibinfo {author} {\bibfnamefont {N.~K.}\ \bibnamefont {Ahmed}}, \bibinfo {author} {\bibfnamefont {A.~F.}\ \bibnamefont {Atiya}}, \bibinfo {author} {\bibfnamefont {N.~E.}\ \bibnamefont {Gayar}},\ and\ \bibinfo {author} {\bibfnamefont {H.}~\bibnamefont {El-Shishiny}},\ }\bibfield  {title} {\bibinfo {title} {An empirical comparison of machine learning models for time series forecasting},\ }\href@noop {} {\bibfield  {journal} {\bibinfo  {journal} {Econometric Reviews}\ }\textbf {\bibinfo {volume} {29}},\ \bibinfo {pages} {594} (\bibinfo {year} {2010})}\BibitemShut {NoStop}%
\bibitem [{\citenamefont {Masini}\ \emph {et~al.}(2023)\citenamefont {Masini}, \citenamefont {Medeiros},\ and\ \citenamefont {Mendes}}]{AItsf2}%
  \BibitemOpen
  \bibfield  {author} {\bibinfo {author} {\bibfnamefont {R.~P.}\ \bibnamefont {Masini}}, \bibinfo {author} {\bibfnamefont {M.~C.}\ \bibnamefont {Medeiros}},\ and\ \bibinfo {author} {\bibfnamefont {E.~F.}\ \bibnamefont {Mendes}},\ }\bibfield  {title} {\bibinfo {title} {Machine learning advances for time series forecasting},\ }\href@noop {} {\bibfield  {journal} {\bibinfo  {journal} {Journal of Economic Surveys}\ }\textbf {\bibinfo {volume} {37}},\ \bibinfo {pages} {76} (\bibinfo {year} {2023})}\BibitemShut {NoStop}%
\bibitem [{\citenamefont {Hajirahimi}\ and\ \citenamefont {Khashei}(2023)}]{AItsf3}%
  \BibitemOpen
  \bibfield  {author} {\bibinfo {author} {\bibfnamefont {Z.}~\bibnamefont {Hajirahimi}}\ and\ \bibinfo {author} {\bibfnamefont {M.}~\bibnamefont {Khashei}},\ }\bibfield  {title} {\bibinfo {title} {Hybridization of hybrid structures for time series forecasting: a review},\ }\href@noop {} {\bibfield  {journal} {\bibinfo  {journal} {Artificial Intelligence Review}\ }\textbf {\bibinfo {volume} {56}},\ \bibinfo {pages} {1201} (\bibinfo {year} {2023})}\BibitemShut {NoStop}%
\bibitem [{\citenamefont {Challu}\ \emph {et~al.}(2023)\citenamefont {Challu}, \citenamefont {Olivares}, \citenamefont {Oreshkin}, \citenamefont {Garza~Ramirez}, \citenamefont {Mergenthaler~Canseco},\ and\ \citenamefont {Dubrawski}}]{AItsf4}%
  \BibitemOpen
  \bibfield  {author} {\bibinfo {author} {\bibfnamefont {C.}~\bibnamefont {Challu}}, \bibinfo {author} {\bibfnamefont {K.~G.}\ \bibnamefont {Olivares}}, \bibinfo {author} {\bibfnamefont {B.~N.}\ \bibnamefont {Oreshkin}}, \bibinfo {author} {\bibfnamefont {F.}~\bibnamefont {Garza~Ramirez}}, \bibinfo {author} {\bibfnamefont {M.}~\bibnamefont {Mergenthaler~Canseco}},\ and\ \bibinfo {author} {\bibfnamefont {A.}~\bibnamefont {Dubrawski}},\ }\bibfield  {title} {\bibinfo {title} {Nhits: Neural hierarchical interpolation for time series forecasting},\ }\href@noop {} {\bibfield  {journal} {\bibinfo  {journal} {Proceedings of the AAAI Conference on Artificial Intelligence}\ }\textbf {\bibinfo {volume} {37}},\ \bibinfo {pages} {6989} (\bibinfo {year} {2023})}\BibitemShut {NoStop}%
\bibitem [{\citenamefont {Orang}\ \emph {et~al.}(2023)\citenamefont {Orang}, \citenamefont {de~Lima~e Silva},\ and\ \citenamefont {Guimarães}}]{AItsf5}%
  \BibitemOpen
  \bibfield  {author} {\bibinfo {author} {\bibfnamefont {O.}~\bibnamefont {Orang}}, \bibinfo {author} {\bibfnamefont {P.~C.}\ \bibnamefont {de~Lima~e Silva}},\ and\ \bibinfo {author} {\bibfnamefont {F.~G.}\ \bibnamefont {Guimarães}},\ }\bibfield  {title} {\bibinfo {title} {Time series forecasting using fuzzy cognitive maps: a survey},\ }\href@noop {} {\bibfield  {journal} {\bibinfo  {journal} {Artificial Intelligence Review}\ }\textbf {\bibinfo {volume} {56}},\ \bibinfo {pages} {7733} (\bibinfo {year} {2023})}\BibitemShut {NoStop}%
\bibitem [{\citenamefont {Wu}\ \emph {et~al.}(2020)\citenamefont {Wu}, \citenamefont {Ding}, \citenamefont {Chan},\ and\ \citenamefont {Chen}}]{AIphy1}%
  \BibitemOpen
  \bibfield  {author} {\bibinfo {author} {\bibfnamefont {B.}~\bibnamefont {Wu}}, \bibinfo {author} {\bibfnamefont {K.}~\bibnamefont {Ding}}, \bibinfo {author} {\bibfnamefont {C.}~\bibnamefont {Chan}},\ and\ \bibinfo {author} {\bibfnamefont {Y.}~\bibnamefont {Chen}},\ }\bibfield  {title} {\bibinfo {title} {Machine prediction of topological transitions in photonic crystals},\ }\href@noop {} {\bibfield  {journal} {\bibinfo  {journal} {Phys. Rev. Appl.}\ }\textbf {\bibinfo {volume} {14}},\ \bibinfo {pages} {044032} (\bibinfo {year} {2020})}\BibitemShut {NoStop}%
\bibitem [{\citenamefont {Ma}\ \emph {et~al.}(2021{\natexlab{a}})\citenamefont {Ma}, \citenamefont {Liu}, \citenamefont {Kudyshev}, \citenamefont {Boltasseva}, \citenamefont {Cai},\ and\ \citenamefont {Liu}}]{AIphy2}%
  \BibitemOpen
  \bibfield  {author} {\bibinfo {author} {\bibfnamefont {W.}~\bibnamefont {Ma}}, \bibinfo {author} {\bibfnamefont {Z.}~\bibnamefont {Liu}}, \bibinfo {author} {\bibfnamefont {Z.~A.}\ \bibnamefont {Kudyshev}}, \bibinfo {author} {\bibfnamefont {A.}~\bibnamefont {Boltasseva}}, \bibinfo {author} {\bibfnamefont {W.}~\bibnamefont {Cai}},\ and\ \bibinfo {author} {\bibfnamefont {Y.}~\bibnamefont {Liu}},\ }\bibfield  {title} {\bibinfo {title} {Deep learning for the design of photonic structures},\ }\href@noop {} {\bibfield  {journal} {\bibinfo  {journal} {Nature Photonics}\ }\textbf {\bibinfo {volume} {15}},\ \bibinfo {pages} {77} (\bibinfo {year} {2021}{\natexlab{a}})}\BibitemShut {NoStop}%
\bibitem [{\citenamefont {Long}\ \emph {et~al.}(2020)\citenamefont {Long}, \citenamefont {Ren},\ and\ \citenamefont {Chen}}]{AIphy3}%
  \BibitemOpen
  \bibfield  {author} {\bibinfo {author} {\bibfnamefont {Y.}~\bibnamefont {Long}}, \bibinfo {author} {\bibfnamefont {J.}~\bibnamefont {Ren}},\ and\ \bibinfo {author} {\bibfnamefont {H.}~\bibnamefont {Chen}},\ }\bibfield  {title} {\bibinfo {title} {Unsupervised manifold clustering of topological phononics},\ }\href@noop {} {\bibfield  {journal} {\bibinfo  {journal} {Phys. Rev. Lett.}\ }\textbf {\bibinfo {volume} {124}},\ \bibinfo {pages} {185501} (\bibinfo {year} {2020})}\BibitemShut {NoStop}%
\bibitem [{\citenamefont {Lu}\ \emph {et~al.}(2020)\citenamefont {Lu}, \citenamefont {Kim},\ and\ \citenamefont {Solja\ifmmode \check{c}\else \v{c}\fi{}i\ifmmode~\acute{c}\else \'{c}\fi{}}}]{AIphy4}%
  \BibitemOpen
  \bibfield  {author} {\bibinfo {author} {\bibfnamefont {P.~Y.}\ \bibnamefont {Lu}}, \bibinfo {author} {\bibfnamefont {S.}~\bibnamefont {Kim}},\ and\ \bibinfo {author} {\bibfnamefont {M.}~\bibnamefont {Solja\ifmmode \check{c}\else \v{c}\fi{}i\ifmmode~\acute{c}\else \'{c}\fi{}}},\ }\bibfield  {title} {\bibinfo {title} {Extracting interpretable physical parameters from spatiotemporal systems using unsupervised learning},\ }\href@noop {} {\bibfield  {journal} {\bibinfo  {journal} {Phys. Rev. X}\ }\textbf {\bibinfo {volume} {10}},\ \bibinfo {pages} {031056} (\bibinfo {year} {2020})}\BibitemShut {NoStop}%
\bibitem [{\citenamefont {Karagiorgi}\ \emph {et~al.}(2022)\citenamefont {Karagiorgi}, \citenamefont {Kasieczka}, \citenamefont {Kravitz}, \citenamefont {Nachman},\ and\ \citenamefont {Shih}}]{AIphy5}%
  \BibitemOpen
  \bibfield  {author} {\bibinfo {author} {\bibfnamefont {G.}~\bibnamefont {Karagiorgi}}, \bibinfo {author} {\bibfnamefont {G.}~\bibnamefont {Kasieczka}}, \bibinfo {author} {\bibfnamefont {S.}~\bibnamefont {Kravitz}}, \bibinfo {author} {\bibfnamefont {B.}~\bibnamefont {Nachman}},\ and\ \bibinfo {author} {\bibfnamefont {D.}~\bibnamefont {Shih}},\ }\bibfield  {title} {\bibinfo {title} {Machine learning in the search for new fundamental physics},\ }\href@noop {} {\bibfield  {journal} {\bibinfo  {journal} {Nature Reviews Physics}\ }\textbf {\bibinfo {volume} {4}},\ \bibinfo {pages} {399} (\bibinfo {year} {2022})}\BibitemShut {NoStop}%
\bibitem [{\citenamefont {Kapp}\ \emph {et~al.}(2022)\citenamefont {Kapp}, \citenamefont {Choi},\ and\ \citenamefont {Hong}}]{AIphy6}%
  \BibitemOpen
  \bibfield  {author} {\bibinfo {author} {\bibfnamefont {S.}~\bibnamefont {Kapp}}, \bibinfo {author} {\bibfnamefont {J.-K.}\ \bibnamefont {Choi}},\ and\ \bibinfo {author} {\bibfnamefont {T.}~\bibnamefont {Hong}},\ }\bibfield  {title} {\bibinfo {title} {Predicting industrial building energy consumption with statistical and machine-learning models informed by physical system parameters},\ }\href@noop {} {\bibfield  {journal} {\bibinfo  {journal} {Renewable and Sustainable Energy Reviews}\ }\textbf {\bibinfo {volume} {172}},\ \bibinfo {pages} {113045} (\bibinfo {year} {2022})}\BibitemShut {NoStop}%
\bibitem [{\citenamefont {Liu}\ \emph {et~al.}(2018)\citenamefont {Liu}, \citenamefont {Tan}, \citenamefont {Khoram},\ and\ \citenamefont {Yu}}]{AIinv1}%
  \BibitemOpen
  \bibfield  {author} {\bibinfo {author} {\bibfnamefont {D.}~\bibnamefont {Liu}}, \bibinfo {author} {\bibfnamefont {Y.}~\bibnamefont {Tan}}, \bibinfo {author} {\bibfnamefont {E.}~\bibnamefont {Khoram}},\ and\ \bibinfo {author} {\bibfnamefont {Z.}~\bibnamefont {Yu}},\ }\bibfield  {title} {\bibinfo {title} {Training deep neural networks for the inverse design of nanophotonic structures},\ }\href@noop {} {\bibfield  {journal} {\bibinfo  {journal} {ACS Photonics}\ }\textbf {\bibinfo {volume} {5}},\ \bibinfo {pages} {1365} (\bibinfo {year} {2018})}\BibitemShut {NoStop}%
\bibitem [{\citenamefont {Peurifoy}\ \emph {et~al.}(2018)\citenamefont {Peurifoy}, \citenamefont {Shen}, \citenamefont {Jing}, \citenamefont {Yang}, \citenamefont {Cano-Renteria}, \citenamefont {DeLacy}, \citenamefont {Joannopoulos}, \citenamefont {Tegmark},\ and\ \citenamefont {Solja{\v{c}}i{\'c}}}]{AIinv2}%
  \BibitemOpen
  \bibfield  {author} {\bibinfo {author} {\bibfnamefont {J.}~\bibnamefont {Peurifoy}}, \bibinfo {author} {\bibfnamefont {Y.}~\bibnamefont {Shen}}, \bibinfo {author} {\bibfnamefont {L.}~\bibnamefont {Jing}}, \bibinfo {author} {\bibfnamefont {Y.}~\bibnamefont {Yang}}, \bibinfo {author} {\bibfnamefont {F.}~\bibnamefont {Cano-Renteria}}, \bibinfo {author} {\bibfnamefont {B.~G.}\ \bibnamefont {DeLacy}}, \bibinfo {author} {\bibfnamefont {J.~D.}\ \bibnamefont {Joannopoulos}}, \bibinfo {author} {\bibfnamefont {M.}~\bibnamefont {Tegmark}},\ and\ \bibinfo {author} {\bibfnamefont {M.}~\bibnamefont {Solja{\v{c}}i{\'c}}},\ }\bibfield  {title} {\bibinfo {title} {Nanophotonic particle simulation and inverse design using artificial neural networks},\ }\href@noop {} {\bibfield  {journal} {\bibinfo  {journal} {Science Advances}\ }\textbf {\bibinfo {volume} {4}},\ \bibinfo {pages} {eaar4206} (\bibinfo {year} {2018})}\BibitemShut {NoStop}%
\bibitem [{\citenamefont {Pilozzi}\ \emph {et~al.}(2018)\citenamefont {Pilozzi}, \citenamefont {Farrelly}, \citenamefont {Marcucci},\ and\ \citenamefont {Conti}}]{AIinv3}%
  \BibitemOpen
  \bibfield  {author} {\bibinfo {author} {\bibfnamefont {L.}~\bibnamefont {Pilozzi}}, \bibinfo {author} {\bibfnamefont {F.~A.}\ \bibnamefont {Farrelly}}, \bibinfo {author} {\bibfnamefont {G.}~\bibnamefont {Marcucci}},\ and\ \bibinfo {author} {\bibfnamefont {C.}~\bibnamefont {Conti}},\ }\bibfield  {title} {\bibinfo {title} {Machine learning inverse problem for topological photonics},\ }\href@noop {} {\bibfield  {journal} {\bibinfo  {journal} {Communications Physics}\ }\textbf {\bibinfo {volume} {1}},\ \bibinfo {pages} {57} (\bibinfo {year} {2018})}\BibitemShut {NoStop}%
\bibitem [{\citenamefont {Long}\ \emph {et~al.}(2019)\citenamefont {Long}, \citenamefont {Ren}, \citenamefont {Li},\ and\ \citenamefont {Chen}}]{AIinv4}%
  \BibitemOpen
  \bibfield  {author} {\bibinfo {author} {\bibfnamefont {Y.}~\bibnamefont {Long}}, \bibinfo {author} {\bibfnamefont {J.}~\bibnamefont {Ren}}, \bibinfo {author} {\bibfnamefont {Y.}~\bibnamefont {Li}},\ and\ \bibinfo {author} {\bibfnamefont {H.}~\bibnamefont {Chen}},\ }\bibfield  {title} {\bibinfo {title} {Inverse design of photonic topological state via machine learning},\ }\href@noop {} {\bibfield  {journal} {\bibinfo  {journal} {Applied Physics Letters}\ }\textbf {\bibinfo {volume} {114}},\ \bibinfo {pages} {181105} (\bibinfo {year} {2019})}\BibitemShut {NoStop}%
\bibitem [{\citenamefont {Jin}\ \emph {et~al.}(2020)\citenamefont {Jin}, \citenamefont {Li}, \citenamefont {Orenstein},\ and\ \citenamefont {Fan}}]{AIinv5}%
  \BibitemOpen
  \bibfield  {author} {\bibinfo {author} {\bibfnamefont {W.}~\bibnamefont {Jin}}, \bibinfo {author} {\bibfnamefont {W.}~\bibnamefont {Li}}, \bibinfo {author} {\bibfnamefont {M.}~\bibnamefont {Orenstein}},\ and\ \bibinfo {author} {\bibfnamefont {S.}~\bibnamefont {Fan}},\ }\bibfield  {title} {\bibinfo {title} {Inverse design of lightweight broadband reflector for relativistic lightsail propulsion},\ }\href@noop {} {\bibfield  {journal} {\bibinfo  {journal} {ACS Photonics}\ }\textbf {\bibinfo {volume} {7}},\ \bibinfo {pages} {2350} (\bibinfo {year} {2020})}\BibitemShut {NoStop}%
\bibitem [{\citenamefont {Guan}\ \emph {et~al.}(2023)\citenamefont {Guan}, \citenamefont {Raza}, \citenamefont {Mao}, \citenamefont {Vega},\ and\ \citenamefont {Zhang}}]{AIinv6}%
  \BibitemOpen
  \bibfield  {author} {\bibinfo {author} {\bibfnamefont {Q.}~\bibnamefont {Guan}}, \bibinfo {author} {\bibfnamefont {A.}~\bibnamefont {Raza}}, \bibinfo {author} {\bibfnamefont {S.~S.}\ \bibnamefont {Mao}}, \bibinfo {author} {\bibfnamefont {L.~F.}\ \bibnamefont {Vega}},\ and\ \bibinfo {author} {\bibfnamefont {T.}~\bibnamefont {Zhang}},\ }\bibfield  {title} {\bibinfo {title} {Machine learning-enabled inverse design of radiative cooling film with on-demand transmissive color},\ }\href {https://doi.org/10.1021/acsphotonics.2c01857} {\bibfield  {journal} {\bibinfo  {journal} {ACS Photonics}\ }\textbf {\bibinfo {volume} {10}},\ \bibinfo {pages} {715} (\bibinfo {year} {2023})}\BibitemShut {NoStop}%
\bibitem [{\citenamefont {Challapalli}\ \emph {et~al.}(2023)\citenamefont {Challapalli}, \citenamefont {Konlan},\ and\ \citenamefont {Li}}]{AIinv7}%
  \BibitemOpen
  \bibfield  {author} {\bibinfo {author} {\bibfnamefont {A.}~\bibnamefont {Challapalli}}, \bibinfo {author} {\bibfnamefont {J.}~\bibnamefont {Konlan}},\ and\ \bibinfo {author} {\bibfnamefont {G.}~\bibnamefont {Li}},\ }\bibfield  {title} {\bibinfo {title} {Inverse machine learning discovered metamaterials with record high recovery stress},\ }\href@noop {} {\bibfield  {journal} {\bibinfo  {journal} {International Journal of Mechanical Sciences}\ }\textbf {\bibinfo {volume} {244}},\ \bibinfo {pages} {108029} (\bibinfo {year} {2023})}\BibitemShut {NoStop}%
\bibitem [{\citenamefont {Ding}\ \emph {et~al.}(2023)\citenamefont {Ding}, \citenamefont {Su}, \citenamefont {Luo}, \citenamefont {Ye}, \citenamefont {Wu},\ and\ \citenamefont {Yao}}]{AIinv8}%
  \BibitemOpen
  \bibfield  {author} {\bibinfo {author} {\bibfnamefont {Z.}~\bibnamefont {Ding}}, \bibinfo {author} {\bibfnamefont {W.}~\bibnamefont {Su}}, \bibinfo {author} {\bibfnamefont {Y.}~\bibnamefont {Luo}}, \bibinfo {author} {\bibfnamefont {L.}~\bibnamefont {Ye}}, \bibinfo {author} {\bibfnamefont {H.}~\bibnamefont {Wu}},\ and\ \bibinfo {author} {\bibfnamefont {H.}~\bibnamefont {Yao}},\ }\bibfield  {title} {\bibinfo {title} {Machine learning in design of broadband terahertz absorbers based on composite structures},\ }\href@noop {} {\bibfield  {journal} {\bibinfo  {journal} {Materials \& Design}\ }\textbf {\bibinfo {volume} {233}},\ \bibinfo {pages} {112215} (\bibinfo {year} {2023})}\BibitemShut {NoStop}%
\bibitem [{\citenamefont {So}\ \emph {et~al.}(2021)\citenamefont {So}, \citenamefont {Yang}, \citenamefont {Lee},\ and\ \citenamefont {Rho}}]{rec1}%
  \BibitemOpen
  \bibfield  {author} {\bibinfo {author} {\bibfnamefont {S.}~\bibnamefont {So}}, \bibinfo {author} {\bibfnamefont {Y.}~\bibnamefont {Yang}}, \bibinfo {author} {\bibfnamefont {T.}~\bibnamefont {Lee}},\ and\ \bibinfo {author} {\bibfnamefont {J.}~\bibnamefont {Rho}},\ }\bibfield  {title} {\bibinfo {title} {On-demand design of spectrally sensitive multiband absorbers using an artificial neural network},\ }\href@noop {} {\bibfield  {journal} {\bibinfo  {journal} {Photonics Research}\ }\textbf {\bibinfo {volume} {9}},\ \bibinfo {pages} {B153} (\bibinfo {year} {2021})}\BibitemShut {NoStop}%
\bibitem [{\citenamefont {Chen}\ \emph {et~al.}(2021)\citenamefont {Chen}, \citenamefont {Ding}, \citenamefont {Li}, \citenamefont {Xi}, \citenamefont {Ye}, \citenamefont {Wu},\ and\ \citenamefont {Wu}}]{rec2}%
  \BibitemOpen
  \bibfield  {author} {\bibinfo {author} {\bibfnamefont {J.}~\bibnamefont {Chen}}, \bibinfo {author} {\bibfnamefont {W.}~\bibnamefont {Ding}}, \bibinfo {author} {\bibfnamefont {X.-M.}\ \bibnamefont {Li}}, \bibinfo {author} {\bibfnamefont {X.}~\bibnamefont {Xi}}, \bibinfo {author} {\bibfnamefont {K.-P.}\ \bibnamefont {Ye}}, \bibinfo {author} {\bibfnamefont {H.-B.}\ \bibnamefont {Wu}},\ and\ \bibinfo {author} {\bibfnamefont {R.-X.}\ \bibnamefont {Wu}},\ }\bibfield  {title} {\bibinfo {title} {Absorption and diffusion enabled ultrathin broadband metamaterial absorber designed by deep neural network and pso},\ }\href {https://doi.org/10.1109/LAWP.2021.3101703} {\bibfield  {journal} {\bibinfo  {journal} {IEEE Antennas and Wireless Propagation Letters}\ }\textbf {\bibinfo {volume} {20}},\ \bibinfo {pages} {1993} (\bibinfo {year} {2021})}\BibitemShut {NoStop}%
\bibitem [{\citenamefont {Hou}\ \emph {et~al.}(2020)\citenamefont {Hou}, \citenamefont {Lin}, \citenamefont {Xu}, \citenamefont {Tian}, \citenamefont {Wang}, \citenamefont {Shi}, \citenamefont {Deng},\ and\ \citenamefont {Chen}}]{rec3}%
  \BibitemOpen
  \bibfield  {author} {\bibinfo {author} {\bibfnamefont {J.}~\bibnamefont {Hou}}, \bibinfo {author} {\bibfnamefont {H.}~\bibnamefont {Lin}}, \bibinfo {author} {\bibfnamefont {W.}~\bibnamefont {Xu}}, \bibinfo {author} {\bibfnamefont {Y.}~\bibnamefont {Tian}}, \bibinfo {author} {\bibfnamefont {Y.}~\bibnamefont {Wang}}, \bibinfo {author} {\bibfnamefont {X.}~\bibnamefont {Shi}}, \bibinfo {author} {\bibfnamefont {F.}~\bibnamefont {Deng}},\ and\ \bibinfo {author} {\bibfnamefont {L.}~\bibnamefont {Chen}},\ }\bibfield  {title} {\bibinfo {title} {Customized inverse design of metamaterial absorber based on target-driven deep learning method},\ }\href {https://doi.org/10.1109/ACCESS.2020.3038933} {\bibfield  {journal} {\bibinfo  {journal} {IEEE Access}\ }\textbf {\bibinfo {volume} {8}},\ \bibinfo {pages} {211849} (\bibinfo {year} {2020})}\BibitemShut {NoStop}%
\bibitem [{\citenamefont {Ma}\ \emph {et~al.}(2020)\citenamefont {Ma}, \citenamefont {Huang}, \citenamefont {Pu}, \citenamefont {Xu}, \citenamefont {Luo}, \citenamefont {Guo},\ and\ \citenamefont {Luo}}]{rec4}%
  \BibitemOpen
  \bibfield  {author} {\bibinfo {author} {\bibfnamefont {J.}~\bibnamefont {Ma}}, \bibinfo {author} {\bibfnamefont {Y.}~\bibnamefont {Huang}}, \bibinfo {author} {\bibfnamefont {M.}~\bibnamefont {Pu}}, \bibinfo {author} {\bibfnamefont {D.}~\bibnamefont {Xu}}, \bibinfo {author} {\bibfnamefont {J.}~\bibnamefont {Luo}}, \bibinfo {author} {\bibfnamefont {Y.}~\bibnamefont {Guo}},\ and\ \bibinfo {author} {\bibfnamefont {X.}~\bibnamefont {Luo}},\ }\bibfield  {title} {\bibinfo {title} {Inverse design of broadband metasurface absorber based on convolutional autoencoder network and inverse design network},\ }\href@noop {} {\bibfield  {journal} {\bibinfo  {journal} {Journal of Physics D: Applied Physics}\ }\textbf {\bibinfo {volume} {53}},\ \bibinfo {pages} {464002} (\bibinfo {year} {2020})}\BibitemShut {NoStop}%
\bibitem [{\citenamefont {Gahlmann}\ and\ \citenamefont {Tassin}(2022)}]{rec5}%
  \BibitemOpen
  \bibfield  {author} {\bibinfo {author} {\bibfnamefont {T.}~\bibnamefont {Gahlmann}}\ and\ \bibinfo {author} {\bibfnamefont {P.}~\bibnamefont {Tassin}},\ }\bibfield  {title} {\bibinfo {title} {Deep neural networks for the prediction of the optical properties and the free-form inverse design of metamaterials},\ }\href {https://doi.org/10.1103/PhysRevB.106.085408} {\bibfield  {journal} {\bibinfo  {journal} {Phys. Rev. B}\ }\textbf {\bibinfo {volume} {106}},\ \bibinfo {pages} {085408} (\bibinfo {year} {2022})}\BibitemShut {NoStop}%
\bibitem [{\citenamefont {Unni}\ \emph {et~al.}(2020)\citenamefont {Unni}, \citenamefont {Yao},\ and\ \citenamefont {Zheng}}]{rec6}%
  \BibitemOpen
  \bibfield  {author} {\bibinfo {author} {\bibfnamefont {R.}~\bibnamefont {Unni}}, \bibinfo {author} {\bibfnamefont {K.}~\bibnamefont {Yao}},\ and\ \bibinfo {author} {\bibfnamefont {Y.}~\bibnamefont {Zheng}},\ }\bibfield  {title} {\bibinfo {title} {Deep convolutional mixture density network for inverse design of layered photonic structures},\ }\href {https://doi.org/10.1021/acsphotonics.0c00630} {\bibfield  {journal} {\bibinfo  {journal} {ACS Photonics}\ }\textbf {\bibinfo {volume} {7}},\ \bibinfo {pages} {2703} (\bibinfo {year} {2020})}\BibitemShut {NoStop}%
\bibitem [{\citenamefont {So}\ \emph {et~al.}(2020)\citenamefont {So}, \citenamefont {Badloe}, \citenamefont {Noh}, \citenamefont {Bravo-Abad},\ and\ \citenamefont {Rho}}]{rec7}%
  \BibitemOpen
  \bibfield  {author} {\bibinfo {author} {\bibfnamefont {S.}~\bibnamefont {So}}, \bibinfo {author} {\bibfnamefont {T.}~\bibnamefont {Badloe}}, \bibinfo {author} {\bibfnamefont {J.}~\bibnamefont {Noh}}, \bibinfo {author} {\bibfnamefont {J.}~\bibnamefont {Bravo-Abad}},\ and\ \bibinfo {author} {\bibfnamefont {J.}~\bibnamefont {Rho}},\ }\bibfield  {title} {\bibinfo {title} {Deep learning enabled inverse design in nanophotonics},\ }\href {https://doi.org/doi:10.1515/nanoph-2019-0474} {\bibfield  {journal} {\bibinfo  {journal} {Nanophotonics}\ }\textbf {\bibinfo {volume} {9}},\ \bibinfo {pages} {1041} (\bibinfo {year} {2020})}\BibitemShut {NoStop}%
\bibitem [{\citenamefont {Kiarashinejad}\ \emph {et~al.}(2020)\citenamefont {Kiarashinejad}, \citenamefont {Abdollahramezani},\ and\ \citenamefont {Adibi}}]{rec9}%
  \BibitemOpen
  \bibfield  {author} {\bibinfo {author} {\bibfnamefont {Y.}~\bibnamefont {Kiarashinejad}}, \bibinfo {author} {\bibfnamefont {S.}~\bibnamefont {Abdollahramezani}},\ and\ \bibinfo {author} {\bibfnamefont {A.}~\bibnamefont {Adibi}},\ }\bibfield  {title} {\bibinfo {title} {Deep learning approach based on dimensionality reduction for designing electromagnetic nanostructures},\ }\href@noop {} {\bibfield  {journal} {\bibinfo  {journal} {npj Computational Materials}\ }\textbf {\bibinfo {volume} {6}},\ \bibinfo {pages} {12} (\bibinfo {year} {2020})}\BibitemShut {NoStop}%
\bibitem [{\citenamefont {An}\ \emph {et~al.}(2021)\citenamefont {An}, \citenamefont {Zheng}, \citenamefont {Tang}, \citenamefont {Shalaginov}, \citenamefont {Zhou}, \citenamefont {Li}, \citenamefont {Kang}, \citenamefont {Richardson}, \citenamefont {Gu}, \citenamefont {Hu}, \citenamefont {Fowler},\ and\ \citenamefont {Zhang}}]{rec10}%
  \BibitemOpen
  \bibfield  {author} {\bibinfo {author} {\bibfnamefont {S.}~\bibnamefont {An}}, \bibinfo {author} {\bibfnamefont {B.}~\bibnamefont {Zheng}}, \bibinfo {author} {\bibfnamefont {H.}~\bibnamefont {Tang}}, \bibinfo {author} {\bibfnamefont {M.~Y.}\ \bibnamefont {Shalaginov}}, \bibinfo {author} {\bibfnamefont {L.}~\bibnamefont {Zhou}}, \bibinfo {author} {\bibfnamefont {H.}~\bibnamefont {Li}}, \bibinfo {author} {\bibfnamefont {M.}~\bibnamefont {Kang}}, \bibinfo {author} {\bibfnamefont {K.~A.}\ \bibnamefont {Richardson}}, \bibinfo {author} {\bibfnamefont {T.}~\bibnamefont {Gu}}, \bibinfo {author} {\bibfnamefont {J.}~\bibnamefont {Hu}}, \bibinfo {author} {\bibfnamefont {C.}~\bibnamefont {Fowler}},\ and\ \bibinfo {author} {\bibfnamefont {H.}~\bibnamefont {Zhang}},\ }\bibfield  {title} {\bibinfo {title} {Multifunctional metasurface design with a generative adversarial network},\ }\href@noop {} {\bibfield  {journal} {\bibinfo  {journal} {Advanced Optical Materials}\ }\textbf {\bibinfo {volume} {9}},\ \bibinfo {pages}
  {2001433} (\bibinfo {year} {2021})}\BibitemShut {NoStop}%
\bibitem [{\citenamefont {Ghosh}\ \emph {et~al.}(2016)\citenamefont {Ghosh}, \citenamefont {Bhattacharyya},\ and\ \citenamefont {Srivastava}}]{LayAbso1}%
  \BibitemOpen
  \bibfield  {author} {\bibinfo {author} {\bibfnamefont {S.}~\bibnamefont {Ghosh}}, \bibinfo {author} {\bibfnamefont {S.}~\bibnamefont {Bhattacharyya}},\ and\ \bibinfo {author} {\bibfnamefont {K.~V.}\ \bibnamefont {Srivastava}},\ }\bibfield  {title} {\bibinfo {title} {Design, characterisation and fabrication of a broadband polarisation-insensitive multi-layer circuit analogue absorber},\ }\href@noop {} {\bibfield  {journal} {\bibinfo  {journal} {IET Microwaves, Antennas \& Propagation}\ }\textbf {\bibinfo {volume} {10}},\ \bibinfo {pages} {850} (\bibinfo {year} {2016})}\BibitemShut {NoStop}%
\bibitem [{\citenamefont {Xiong}\ \emph {et~al.}(2013)\citenamefont {Xiong}, \citenamefont {Hong}, \citenamefont {Luo},\ and\ \citenamefont {Zhong}}]{LayAbso2}%
  \BibitemOpen
  \bibfield  {author} {\bibinfo {author} {\bibfnamefont {H.}~\bibnamefont {Xiong}}, \bibinfo {author} {\bibfnamefont {J.-S.}\ \bibnamefont {Hong}}, \bibinfo {author} {\bibfnamefont {C.-M.}\ \bibnamefont {Luo}},\ and\ \bibinfo {author} {\bibfnamefont {L.-L.}\ \bibnamefont {Zhong}},\ }\bibfield  {title} {\bibinfo {title} {An ultrathin and broadband metamaterial absorber using multi-layer structures},\ }\href {https://doi.org/10.1063/1.4818318} {\bibfield  {journal} {\bibinfo  {journal} {Journal of Applied Physics}\ }\textbf {\bibinfo {volume} {114}},\ \bibinfo {pages} {064109} (\bibinfo {year} {2013})}\BibitemShut {NoStop}%
\bibitem [{\citenamefont {Rozanov}(2000)}]{rozanov2000ultimate}%
  \BibitemOpen
  \bibfield  {author} {\bibinfo {author} {\bibfnamefont {K.}~\bibnamefont {Rozanov}},\ }\bibfield  {title} {\bibinfo {title} {Ultimate thickness to bandwidth ratio of radar absorbers},\ }\href@noop {} {\bibfield  {journal} {\bibinfo  {journal} {IEEE Transactions on Antennas and Propagation}\ }\textbf {\bibinfo {volume} {48}},\ \bibinfo {pages} {1230} (\bibinfo {year} {2000})}\BibitemShut {NoStop}%
\bibitem [{\citenamefont {Kazem~Zadeh}\ and\ \citenamefont {Karlsson}(2009)}]{CapaCir1}%
  \BibitemOpen
  \bibfield  {author} {\bibinfo {author} {\bibfnamefont {A.}~\bibnamefont {Kazem~Zadeh}}\ and\ \bibinfo {author} {\bibfnamefont {A.}~\bibnamefont {Karlsson}},\ }\bibfield  {title} {\bibinfo {title} {Capacitive circuit method for fast and efficient design of wideband radar absorbers},\ }\href {https://doi.org/10.1109/TAP.2009.2024490} {\bibfield  {journal} {\bibinfo  {journal} {IEEE Transactions on Antennas and Propagation}\ }\textbf {\bibinfo {volume} {57}},\ \bibinfo {pages} {2307} (\bibinfo {year} {2009})}\BibitemShut {NoStop}%
\bibitem [{\citenamefont {Costa}\ \emph {et~al.}(2010)\citenamefont {Costa}, \citenamefont {Monorchio},\ and\ \citenamefont {Manara}}]{CapaCir2}%
  \BibitemOpen
  \bibfield  {author} {\bibinfo {author} {\bibfnamefont {F.}~\bibnamefont {Costa}}, \bibinfo {author} {\bibfnamefont {A.}~\bibnamefont {Monorchio}},\ and\ \bibinfo {author} {\bibfnamefont {G.}~\bibnamefont {Manara}},\ }\bibfield  {title} {\bibinfo {title} {Analysis and design of ultra thin electromagnetic absorbers comprising resistively loaded high impedance surfaces},\ }\href {https://doi.org/10.1109/TAP.2010.2044329} {\bibfield  {journal} {\bibinfo  {journal} {IEEE Transactions on Antennas and Propagation}\ }\textbf {\bibinfo {volume} {58}},\ \bibinfo {pages} {1551} (\bibinfo {year} {2010})}\BibitemShut {NoStop}%
\bibitem [{\citenamefont {Costa}\ \emph {et~al.}(2009)\citenamefont {Costa}, \citenamefont {Monorchio},\ and\ \citenamefont {Manara}}]{CapaCir3}%
  \BibitemOpen
  \bibfield  {author} {\bibinfo {author} {\bibfnamefont {F.}~\bibnamefont {Costa}}, \bibinfo {author} {\bibfnamefont {A.}~\bibnamefont {Monorchio}},\ and\ \bibinfo {author} {\bibfnamefont {G.}~\bibnamefont {Manara}},\ }\bibfield  {title} {\bibinfo {title} {An equivalent circuit model of frequency selective surfaces embedded within dielectric layers},\ }in\ \href {https://doi.org/10.1109/APS.2009.5171774} {\emph {\bibinfo {booktitle} {2009 IEEE Antennas and Propagation Society International Symposium}}}\ (\bibinfo {year} {2009})\ pp.\ \bibinfo {pages} {1--4}\BibitemShut {NoStop}%
\bibitem [{\citenamefont {{Langley}}\ and\ \citenamefont {{Parker}}(1982)}]{RCmodel}%
  \BibitemOpen
  \bibfield  {author} {\bibinfo {author} {\bibfnamefont {R.~J.}\ \bibnamefont {{Langley}}}\ and\ \bibinfo {author} {\bibfnamefont {E.~A.}\ \bibnamefont {{Parker}}},\ }\bibfield  {title} {\bibinfo {title} {{Equivalent circuit model for arrays of square loops}},\ }\href@noop {} {\bibfield  {journal} {\bibinfo  {journal} {Electronics Letters}\ }\textbf {\bibinfo {volume} {18}},\ \bibinfo {pages} {294} (\bibinfo {year} {1982})}\BibitemShut {NoStop}%
\bibitem [{\citenamefont {Pozar}(2011)}]{CapaCir4}%
  \BibitemOpen
  \bibfield  {author} {\bibinfo {author} {\bibfnamefont {D.~M.}\ \bibnamefont {Pozar}},\ }\href@noop {} {\emph {\bibinfo {title} {Microwave engineering}}}\ (\bibinfo  {publisher} {John wiley \& sons},\ \bibinfo {year} {2011})\BibitemShut {NoStop}%
\bibitem [{\citenamefont {Hornik}(1991)}]{UnivAppr}%
  \BibitemOpen
  \bibfield  {author} {\bibinfo {author} {\bibfnamefont {K.}~\bibnamefont {Hornik}},\ }\bibfield  {title} {\bibinfo {title} {Approximation capabilities of multilayer feedforward networks},\ }\href@noop {} {\bibfield  {journal} {\bibinfo  {journal} {Neural Networks}\ }\textbf {\bibinfo {volume} {4}},\ \bibinfo {pages} {251} (\bibinfo {year} {1991})}\BibitemShut {NoStop}%
\bibitem [{\citenamefont {Zhang}\ \emph {et~al.}(2019)\citenamefont {Zhang}, \citenamefont {Yang}, \citenamefont {Cao}, \citenamefont {Yuan}, \citenamefont {Ke}, \citenamefont {Yang}, \citenamefont {Cheng},\ and\ \citenamefont {Cui}}]{ITO}%
  \BibitemOpen
  \bibfield  {author} {\bibinfo {author} {\bibfnamefont {C.}~\bibnamefont {Zhang}}, \bibinfo {author} {\bibfnamefont {J.}~\bibnamefont {Yang}}, \bibinfo {author} {\bibfnamefont {W.}~\bibnamefont {Cao}}, \bibinfo {author} {\bibfnamefont {W.}~\bibnamefont {Yuan}}, \bibinfo {author} {\bibfnamefont {J.}~\bibnamefont {Ke}}, \bibinfo {author} {\bibfnamefont {L.}~\bibnamefont {Yang}}, \bibinfo {author} {\bibfnamefont {Q.}~\bibnamefont {Cheng}},\ and\ \bibinfo {author} {\bibfnamefont {T.}~\bibnamefont {Cui}},\ }\bibfield  {title} {\bibinfo {title} {Transparently curved metamaterial with broadband millimeter wave absorption},\ }\href@noop {} {\bibfield  {journal} {\bibinfo  {journal} {Photonics Research}\ }\textbf {\bibinfo {volume} {7}},\ \bibinfo {pages} {478} (\bibinfo {year} {2019})}\BibitemShut {NoStop}%
\bibitem [{\citenamefont {Munk}(2005)}]{Cir2}%
  \BibitemOpen
  \bibfield  {author} {\bibinfo {author} {\bibfnamefont {B.~A.}\ \bibnamefont {Munk}},\ }\href {https://doi.org/https://doi.org/10.1002/0471723770.fmatter} {\emph {\bibinfo {title} {Frequency selective surfaces: theory and design}}}\ (\bibinfo  {publisher} {John Wiley \& Sons},\ \bibinfo {year} {2005})\BibitemShut {NoStop}%
\bibitem [{\citenamefont {Munk}(2009)}]{Cir3}%
  \BibitemOpen
  \bibfield  {author} {\bibinfo {author} {\bibfnamefont {B.~A.}\ \bibnamefont {Munk}},\ }\href@noop {} {\emph {\bibinfo {title} {Metamaterials: critique and alternatives}}}\ (\bibinfo  {publisher} {John Wiley \& Sons},\ \bibinfo {year} {2009})\BibitemShut {NoStop}%
\bibitem [{\citenamefont {Munk}\ \emph {et~al.}(2007)\citenamefont {Munk}, \citenamefont {Munk},\ and\ \citenamefont {Pryor}}]{Cir4}%
  \BibitemOpen
  \bibfield  {author} {\bibinfo {author} {\bibfnamefont {B.~A.}\ \bibnamefont {Munk}}, \bibinfo {author} {\bibfnamefont {P.}~\bibnamefont {Munk}},\ and\ \bibinfo {author} {\bibfnamefont {J.}~\bibnamefont {Pryor}},\ }\bibfield  {title} {\bibinfo {title} {On designing jaumann and circuit analog absorbers (ca absorbers) for oblique angle of incidence},\ }\href@noop {} {\bibfield  {journal} {\bibinfo  {journal} {IEEE Transactions on Antennas and Propagation}\ }\textbf {\bibinfo {volume} {55}},\ \bibinfo {pages} {186} (\bibinfo {year} {2007})}\BibitemShut {NoStop}%
\bibitem [{\citenamefont {Sun}\ \emph {et~al.}(2011)\citenamefont {Sun}, \citenamefont {Liu}, \citenamefont {Dong},\ and\ \citenamefont {Zhou}}]{LayAbso3}%
  \BibitemOpen
  \bibfield  {author} {\bibinfo {author} {\bibfnamefont {J.}~\bibnamefont {Sun}}, \bibinfo {author} {\bibfnamefont {L.}~\bibnamefont {Liu}}, \bibinfo {author} {\bibfnamefont {G.}~\bibnamefont {Dong}},\ and\ \bibinfo {author} {\bibfnamefont {J.}~\bibnamefont {Zhou}},\ }\bibfield  {title} {\bibinfo {title} {An extremely broad band metamaterial absorber based on destructive interference},\ }\href {https://doi.org/10.1364/OE.19.021155} {\bibfield  {journal} {\bibinfo  {journal} {Optical Express}\ }\textbf {\bibinfo {volume} {19}},\ \bibinfo {pages} {21155} (\bibinfo {year} {2011})}\BibitemShut {NoStop}%
\bibitem [{\citenamefont {Yang}\ \emph {et~al.}(2017)\citenamefont {Yang}, \citenamefont {Chen}, \citenamefont {Fu},\ and\ \citenamefont {Sheng}}]{OptimalAbso}%
  \BibitemOpen
  \bibfield  {author} {\bibinfo {author} {\bibfnamefont {M.}~\bibnamefont {Yang}}, \bibinfo {author} {\bibfnamefont {S.}~\bibnamefont {Chen}}, \bibinfo {author} {\bibfnamefont {C.}~\bibnamefont {Fu}},\ and\ \bibinfo {author} {\bibfnamefont {P.}~\bibnamefont {Sheng}},\ }\bibfield  {title} {\bibinfo {title} {Optimal sound-absorbing structures},\ }\href {https://doi.org/10.1039/C7MH00129K} {\bibfield  {journal} {\bibinfo  {journal} {Material Horizon}\ }\textbf {\bibinfo {volume} {4}},\ \bibinfo {pages} {673} (\bibinfo {year} {2017})}\BibitemShut {NoStop}%
\bibitem [{\citenamefont {Sch{\"o}lkopf}\ and\ \citenamefont {Smola}(2002)}]{Reg1}%
  \BibitemOpen
  \bibfield  {author} {\bibinfo {author} {\bibfnamefont {B.}~\bibnamefont {Sch{\"o}lkopf}}\ and\ \bibinfo {author} {\bibfnamefont {A.~J.}\ \bibnamefont {Smola}},\ }\href@noop {} {\emph {\bibinfo {title} {Learning with kernels: support vector machines, regularization, optimization, and beyond}}}\ (\bibinfo  {publisher} {MIT press},\ \bibinfo {year} {2002})\BibitemShut {NoStop}%
\bibitem [{\citenamefont {Ma}\ \emph {et~al.}(2021{\natexlab{b}})\citenamefont {Ma}, \citenamefont {Liu}, \citenamefont {Kudyshev}, \citenamefont {Boltasseva}, \citenamefont {Cai},\ and\ \citenamefont {Liu}}]{rec8}%
  \BibitemOpen
  \bibfield  {author} {\bibinfo {author} {\bibfnamefont {W.}~\bibnamefont {Ma}}, \bibinfo {author} {\bibfnamefont {Z.}~\bibnamefont {Liu}}, \bibinfo {author} {\bibfnamefont {Z.~A.}\ \bibnamefont {Kudyshev}}, \bibinfo {author} {\bibfnamefont {A.}~\bibnamefont {Boltasseva}}, \bibinfo {author} {\bibfnamefont {W.}~\bibnamefont {Cai}},\ and\ \bibinfo {author} {\bibfnamefont {Y.}~\bibnamefont {Liu}},\ }\bibfield  {title} {\bibinfo {title} {Deep learning for the design of photonic structures},\ }\href {https://doi.org/10.1038/s41566-020-0685-y} {\bibfield  {journal} {\bibinfo  {journal} {Nature Photonics}\ }\textbf {\bibinfo {volume} {15}},\ \bibinfo {pages} {77} (\bibinfo {year} {2021}{\natexlab{b}})}\BibitemShut {NoStop}%
\end{thebibliography}%

\end{document}


\title{Constrained tandem neural network assisted inverse design of metasurfaces for microwave absorption: supplemental document}
\maketitle

\begin{figure}[ht]
\centering
\includegraphics[width=\textwidth]{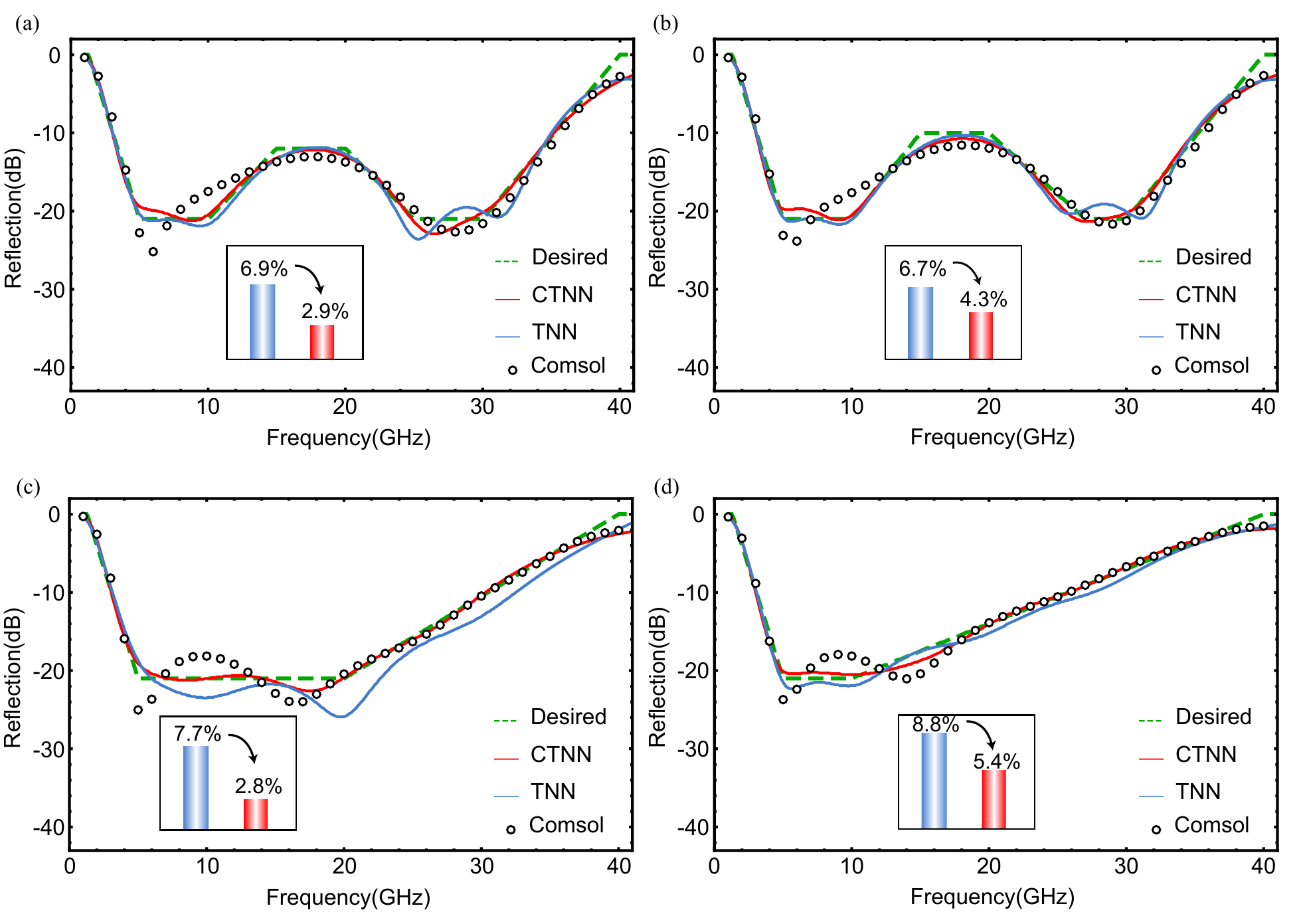}
\caption{More designs the CTNN gives. (a) and (b) are the "W" tyep absorbers. (c) and (d) are absorbers designed for a specific frequency band.}
\label{FIG1}
\end{figure}

\begin{table}[ht]
\centering
\caption{The design parameters of the examples in Fig. \ref{FIG1}.}
\begin{ruledtabular}
\begin{tabular}{  p{2.5cm}  p{2.5cm}  p{2.5cm}  p{2.5cm}  p{2.5cm}}
Parameter & Fig. \ref{FIG1}(a) CTNN & Fig. \ref{FIG1}(a) TNN & Fig. \ref{FIG1}(b) CTNN & Fig. \ref{FIG1}(b) TNN \\
\hline
$\epsilon_S$ & 1.20  & 1.20             & 1.20            & 1.20            \\
$\epsilon_C$ & 6.81  & 7.44             & 6.76          & 6.67          \\
$T_C$(mm)    & 0.127 & 0.103            & 0.143        & 0.122          \\
$d_1$(mm)    & 3.40  & 3.74             & 3.46          & 3.73          \\
$d_2$(mm)    & 2.50  & 2.57             & 2.46          & 2.48          \\
$d_3$(mm)    & 3.37  & 3.51             & 3.32          & 3.45          \\
$d_4$(mm)    & 2.34  & 2.35             & 2.28          & 2.23          \\
$w_1$(mm)    & 6.42  & 6.47             & 6.36            & 6.41            \\
$w_2$(mm)    & 5.98  & 5.80             & 6.00            & 5.80            \\
$w_3$(mm)    & 2.99  & 3.06             & 3.05            & 3.08            \\
$w_4$(mm)    & 1.17  & 1.06             & 1.19            & 1.08           \\
$a_1$(mm)    & 6.80  & 6.80             & 6.80            & 6.80            \\
$a_2$(mm)    & 6.80  & 6.80             & 6.80            & 6.80            \\
$a_3$(mm)    & 3.40  & 3.40             & 3.40            & 3.40            \\
$a_4$(mm)    & 1.70  & 1.70             & 1.70            & 1.70            \\
$Rs_1$($\Omega$/Sq) & 180   & 213             & 192          & 214          \\
$Rs_2$($\Omega$/Sq) & 306   & 299             & 290          & 283         \\
$Rs_3$($\Omega$/Sq) & 686   & 726             & 791          & 814          \\
$Rs_4$($\Omega$/Sq) & 556   & 432             & 522           & 400      \\
\end{tabular}
\label{table1}
\end{ruledtabular}
\end{table}

\begin{table}[ht]
\centering
\begin{ruledtabular}
\begin{tabular}{  p{2.5cm}  p{2.5cm}  p{2.5cm}  p{2.5cm}  p{2.5cm}}
Parameter & Fig. \ref{FIG1}(c) CTNN & Fig. \ref{FIG1}(c) TNN & Fig. \ref{FIG1}(d) CTNN & Fig. \ref{FIG1}(d) TNN \\
\hline
$\epsilon_S$        & 1.20           & 1.20             & 1.20            & 1.20            \\
$\epsilon_C$         & 6.53      & 8.13             & 6.95          & 8.31          \\
$T_C$(mm)    & 0.115    & 0.0628            & 0.178        & 0.133          \\
$d_1$(mm)    & 2.63  & 3.19             & 2.51          & 2.98          \\
$d_2$(mm)    & 2.92  & 2.96             & 2.76          & 2.85          \\
$d_3$(mm)    & 3.26  & 3.49             & 2.89          & 2.96          \\
$d_4$(mm)    & 3.20  & 3.16             & 3.46          & 3.41          \\
$w_1$(mm)    & 6.68  & 6.42             & 6.59            & 6.40            \\
$w_2$(mm)    & 6.10  & 5.71             & 6.09            & 5.71            \\
$w_3$(mm)    & 3.07 & 3.18            & 3.18            & 3.21            \\
$w_4$(mm)    & 1.16 & 0.986             & 1.25            & 1.11            \\
$a_1$(mm)    & 6.80      & 6.80             & 6.80            & 6.80            \\
$a_2$(mm)    & 6.80       & 6.80             & 6.80            & 6.80            \\
$a_3$(mm)    & 6.80      & 3.40             & 3.40            & 3.40            \\
$a_4$(mm)    & 6.80      & 1.70             & 1.70            & 1.70            \\
$Rs_1$($\Omega$/Sq) & 151   & 167             & 153          & 169          \\
$Rs_2$($\Omega$/Sq) & 307   & 329             & 319          & 313          \\
$Rs_3$($\Omega$/Sq) & 487   & 587             & 526          & 630          \\
$Rs_4$($\Omega$/Sq) & 668   & 525             & 661           & 537      \\
\end{tabular}
\end{ruledtabular}
\end{table}

Fig. \ref{FIG1} shows more designs given by the CTNN. The desired spectra are denoted by the green dashed line, while the absorption response obtained by the
improved CTNN, the traditional TNN, and the full-wave numerical simulation are
represented by the red line, blue line, and open circles, respectively. The parameters of designs are shown in Table \ref{table1}. The inset graphs show the percentage difference between the thickness of the design and the thickness given by the causality limit, showing the CTNN always gives a thinner absorber. The blue (red) bar represents the design given by TNN (CTNN). It is evident that CTNN is more capable of providing designs that approach the causality limit compared to the traditional TNN. 
Besides, these results show that our model is applicable to a wide range of desired spectra.

